\documentclass[%
 reprint,
amsmath,amssymb,
 aps,
floatfix,
]{revtex4-2}

\usepackage{graphicx}
\usepackage{dcolumn}
\usepackage{bm}

\begin{document}

\preprint{APS/123-QED}

\title{Heat Equation from Exact Renormalization Group Equation (ERGE) at Local Potential Approximation (LPA)}

\author{Phumudzo T. Rabambi}
 \homepage{teflon.ac.za@gmail.com}
\affiliation{Department of Physics\\
	University of Witswatersrand\\
	Wits, 2050, South Africa 
}%


\begin{abstract}
\noindent By simply applying the Local Potential Approximation (LPA) on the Polchinski's Exact Renormalization Group (ERG) flow equation for single Bosonic and spinless Fermionic fields, and initially considering only the coarse-graining (blocking) aspect of Wilson's Renormalization Group program. Within the LPA limit the Polchinski's ERG flow equation simplifies into a heat differential equation for both the Bosonic and Fermionic fields. Solving the differential equations leads to logarithmic interactions  (logarithmic vertex function) in both the Bosonic and Fermionic fields at their fixed points.
\end{abstract}

\maketitle


\section{Introduction}

\noindent Within the Framework of Exact Renormalization Group (ERG) methods \cite{PhysRevB.4.3174,article,PhysRevA.8.401,POLCHINSKI1984269}, with a particular focus on Polchinski's  Exact Renormalization Group Equation (ERGE) \cite{POLCHINSKI1984269}, which is the Wilson's Exact Renormalization Group (ERG) \cite{PhysRevB.4.3174} derived in the field theory context. We study a system described by a general effective action consisting of single Bosonic and spinless Fermionic field modes including their interactions. Working in Euclidean momentum space where the theory is regularized by employing the cutoff functions of generic form $K_{UV}(p,\Lambda)$ in both fields, whereby the cutoff functions suppress field modes with momentum $p$ that are above the physical UV (ultra-violet) cutoff scale $\Lambda$. The Polchinski's ERGE (Exact Renormalization Group Equation) emanate from requiring that as the scale $\Lambda$ is lowered by integrating (course-graining) high momentum modes, the effective action (vertex functions) of the theory should be varied in such a way that the partition function of the theory remains unchanged. The variation of the effective action due to lowering of the physical cutoff scale is governed by the Renormalization Group (RG) equation, which is an exact differential equation subsequently giving rise to ERG methods. ERG methods in field theory are exact non-perturbative formulations of RG flow physics encoding it as a differential variation of physical scale and the effective action. They describe the evolution of the renormalized effective action along the RG flow trajectory in terms of differential equations. From the theory setup, these equations arise from the decimation (integration) of fields modes momenta, followed by lowering of the momentum cutoff and then rescaling the action parameters while enforcing the physical observables to remain invariant by requiring the partition function to remain invariant. Usually the RG flow has two physical cutoff scales, one is a high energy scale denoted as the UV cutoff scale, and the second one is a low energy scale denoted as an IR (infrared) cutoff scale, and the RG flow will normally be a flow from the UV (high energy) limit towards the IR (low energy) limit. A generic RG flow equation has the form\\
\\
\begin{align}
\Lambda\frac{\partial}{\partial\Lambda}S_{\Lambda}[\Phi]=\mathcal{F}[S_{\Lambda}],
\end{align}
\noindent where $S_{\Lambda}$ is the effective action and $\Phi$ represents a generic set of fields. An important aspect to solve from the RG flow equations are the fixed points of the theory, which is the scale invariant continuum limit of the theory, and the equation to solve to obtain these fixed points is,
\begin{align}
\Lambda\frac{\partial}{\partial\Lambda}S_{\Lambda}[\Phi]=0.
\end{align}
\noindent Generally solving such an equation specially in the ERGE context can be challenging, since a handful of ERG's problems end up as integro-differential equations \cite{PhysRevB.4.3174,article,PhysRevA.8.401,POLCHINSKI1984269}, with Polchinski's ERG flow equation being in this category. Therefore finding an exact non-perturbative general solution from these kind of equations generally it is a difficult tasks, in practice one ends up with an infinite tower of equations and has to either truncate the equations by performing certain approximations. Several non-perturbative approximation solutions and methods are possible, such as derivative expansion, which involves expanding the equations in powers of momenta and this has shown to yield promising results for calculating critical exponents and renormalization group flows equations in $2\leq d\leq 4$ scalar fields theories \cite{Hasenfratz:1985dm,FILIPPOV1992195,Haagensen_1994,Ball_1995}. The simplest non-perturbative approximation from such power momentum expansion is obtained from taking the leading term (zeroth order term of the expansion), which is the Local Potential Approximation (LPA) expansion. Here we will make use of such power momentum expansion, restrict the computations within the LPA limit. As a start we avoid the consideration of next to leading order terms and other higher order terms in momentum or derivative expansion since these terms are known to yield physical solutions (physical observables) which are scheme dependent (cutoff scale dependent) \cite{Ball_1995,Morris_1994,Tetradis_1994}, which makes it difficult to compute physical observables such as the critical exponents ambiguously. Whilst the advantage with the leading order solutions is that, the LPA physical observables solutions are scheme independent and can be computed unambiguously. \\
\\
It is known and alluded \cite{Cotler_2023} that the Polchinski's ERG flow equation can be written as a functional heat equation. However, presently it appears there is lack of visible available work towards unpacking this alluded notion at a basic level. Clearly the notion that the Polchinski's ERG flow equation can be represented as a functional heat equation hasn't received much attention, and consequently not much work has been done to unpack and elucidate this notion with simple theoretical approaches. Therefore this paper aims to address this, by providing a simple connection or relation between the Polchinski's ERG flow equation and the well known standard heat equation within the LPA limit. Utilizing this notion, in one of our main results in section 3 at equation \eqref{homog} in the case of single Bosonic fields, we show  that at low energy after relaxing the rescaling steps of Wilson's Renormalization Group program and only implement the coarse-graining step, subsequently the Polchinski's ERG flow equation within the LPA limit simplifies into a standard heat equation,
\begin{align}
\frac{\partial}{\partial t}U(z,t)=\frac{\partial^2}{\partial z\partial z}U(z,t),
\end{align}
\\
\noindent where $U(z,t)=e^{-V(z,t)}$, $t$ is a dimensionless parameter, $z$ represent Bosonic fields and $V(z,t)$ is the vertex or interaction function. We will explain more about the results and how they are obtained in later sections of the paper. It is worth mentioning that the results are interesting due to their simplistic nature of relating the Polchinski's ERG flow equation within the LPA limit to a heat equation, this can potentially open a new front to generally study ERG equations through the available technology and language already well developed and established in studying heat equations. Subsequently the solution to this standard heat equation at the fixed point, which is related to the steady state solution in the language of heat equations, we find that the fixed point solution is governed by logarithmic interactions,
\begin{align}
V_*(z,t) = \ln\big(z\sqrt{2\pi}\big)+\frac{1}{2},
\end{align}
\noindent  where $V_*(z,t)$ is the interaction function at the fixed point obtained at $t=\frac{z^2}{2\lambda}$. The solution is interesting since fixed points normally signals a Conformal Field Theory (CFT), therefore at this low energy limit we have established a logarithmic interaction CFT for the single Bosonic fields. Contrary to perturbing the solution in the vicinity of the Gaussian fixed point \cite{Comellas_1998,cmp/1104159469,Bagnuls_2001,Hasenfratz:1985dm}, here we perturb the solution in the vicinity of the logarithmic fixed point obtained above. For the perturbation recipe we use a small parameter to linearize the Polchinski's ERG flow equation in the vicinity of the logarithmic fixed point, but here we implement both rescaling and coarse-graining steps on the Polchinski's ERG flow equation in order to find non-trivial fixed-point solutions. This leads to an infinite set tower of Hermite differentials at each expansion order in parameter $t$, with the Hermite polynomials being solutions to these differential equations. Treating the Fermionic fields in similar fashion, in section 4 at equation \eqref{fermion_results_intro2} we obtain the following analogue of a heat equation in Grassmann variables,
\\
\begin{align}
\frac{\partial}{\partial t} U(z,\bar{z},t)=\frac{\partial^2}{\partial z\partial\bar{z}}U(z,\bar{z},t),
\end{align}
\\
\noindent where $U(z,\bar{z},t)=e^{-V(z,\bar{z},t)}$, $z$ and $\bar{z}$ are Grassmann variables representing Fermionic fields, and $V(z,\bar{z},t)$ is the interaction function. The fixed point solution obtained at $t=\frac{z\bar{z}}{\lambda}$ also yields a logarithmic interaction function,
\\
\begin{align}
V_*\big(z,\bar{z},t\big)=&\big(1+\ln(z\bar{z})\big),
\end{align}
\noindent  which also leads to similar conclusions as in the Bosonic fields case regarding the fixed point being controlled by a logarithmic interaction CFT. However, the nilpotent property in Grassmann variables truncates the expansion in parameter $t$ and renders it finite, which leads to a finite set tower of Grassmann differential equations. 
\\
\\
\noindent The paper is organised as follows. In section 2 we discuss the derivation of Polchinski's ERG flow equation for a system containing both the single Bosonic and spinless Fermionic fields, and only considering the coarse-graining step of Wislon's Renormalization Group program. In section 3 we additionally consider the rescaling step of Wilson's Renormalization Group program on the Polschinki's ERG flow equation, we apply the rescaling step and take the LPA limit of the Polchinski's ERG flow equation, then the differential equation emanating from Polchinki's ERG flow equation at the LPA limit is solved, followed by finding the eigenvalues, eigenfunctions and the critical exponents for the Bosonic fields sector. Then section 4 deals with solving the Polchinski's ERG flow equation within the LPA limit in the Fermionic fields sector.

\section{Derivation of Polchinski's ERGE for Bosonic and Fermionic Fields.}

\noindent The first part of this section provides a review of some useful materials and known results regarding the Polchinski's ERG flow equation. At the end we use these known results to derive results of our own which we will use later on in the paper. Our own final derivation results which we are after and are worth noting are given in equation \eqref{general_polchinski} and equation \eqref{integro}, these results represents Polchinski's ERG flow equation between interacting Bosonic and spinless Fermionic. Readers familiar with the material of this section can just note the mentioned equations above, before deciding to skip this section and move on to the next sections. \\
\\
ERGE equations are derived from two steps which subsequently follow one another repetitively. The first step is a stepwise partial infinitesimal integration of momentum modes, followed by the second step which involves rescaling of variables (operators). In the field theory context, the first step involving the path integral approach has been well elaborated and derived in \cite{POLCHINSKI1984269}, which consequently leads to Polchinski's ERG flow equation, and the second step in field theory context is well covered in \cite{Bagnuls_2001, Rosten_2012, Ball_1995} pertaining to implementing rescaling on Polchinski's ERG flow equation. Here we will derive a similar Polchinki's ERG flow equation for an interacting system of spinless Fermionic and Bosonic fields. For simplicity, we begin by presenting the UV-regularized generating functional of connected Greens function for a simple self-interacting Bosonic field \cite{golner2000exact,Ellwanger_1993,Ellwanger_1994}, and as a start the Fermionic fields will be omitted from the discussion of general features of the flow equation, and they will be included later. We begin by a bare theory regularized in the UV by the cutoff scale $\Lambda_0$ through the propagator, which also includes a running infrared cutoff scale $\Lambda$, and the UV-regularized generating function of connected Greens function will be

\begin{align}
e^{-W_{\Lambda}^{\Lambda_0}[J]}=&\int D\phi e^{-\int_p\frac{1}{2}\phi(p)G^{-1}_B(\frac{p}{\Lambda})\phi(-p)+\int_pJ(p)\phi(-p)-S_{int}^{\Lambda_0}[\phi]},\\\nonumber,
\end{align}

\noindent where on the RHS we have $J(p)$ as the external source field coupled to the Boson field $\phi(p)$, the kinetic part of the theory is represented by the inverse propagator $G_B^{-1}\big(\frac{p}{\Lambda}\big)$ and $S_{int}^{\Lambda_0}[\phi]$ represent the interactions part of the bare theory. Generically $G_B\big(\frac{p}{\Lambda}\big)$ is written as follows
\\
\begin{align}
G_B\bigg(\frac{p}{\Lambda}\bigg)=\frac{K_B\big(\frac{p^2}{\Lambda^2}\big)}{p^2},
\end{align}

\noindent where $K_B\big(\frac{p^2}{\Lambda^2}\big)$ is the UV regulating cutoff function, which is responsible for suppressing high energy momentum modes, the function falls rapidly as $\frac{p^2}{\Lambda^2}\rightarrow\infty$ and it behaves as follows,

\begin{align}
&K_B\bigg(\frac{p^2}{\Lambda^2}\bigg)\rightarrow 1\qquad\text{for}\qquad p^2\ll\Lambda^2\\\nonumber
&K_B\bigg(\frac{p^2}{\Lambda^2}\bigg)\rightarrow 0\qquad\text{for}\qquad p^2\gg\Lambda^2,
\end{align}

\noindent with $\Lambda = \Lambda_0e^{-t}$, $\Lambda_0$ being the UV regulator scale which is held fixed, and parameter $t$ being the label that parametrizes the RG flow and it is dimensionless and usually is a ratio of $\Lambda$ and $\Lambda_0$. Normally the Exact Renormalization Group flow employs two momentum scales, the first scale is the UV regulator scale $\Lambda_0$, which remains fixed during the flow, the second scale is the infrared (IR) flow scale $\Lambda$, which controls the flow down to where high momentum modes have been integrated out, with all modes with $|p|>\Lambda$ being already eliminated (integrated). What is important in the flow is the ratio of both scales $\Lambda_0$ and $\Lambda$, and this ratio can be captured by the dimensional parameter $t=\ln\big(\frac{\Lambda_0}{\Lambda}\big)$. Then during the renormalization group flow starting from the UV ($t=0$, $\Lambda=\Lambda_0$) flowing down to the IR ($t\rightarrow \infty$, $\Lambda\rightarrow 0$), $\Lambda$ is reduced from $\Lambda_0$ to zero while $S_{int}[\phi,\Lambda]$ evolves in such a way that the UV regularized generating function remains invariant, and this is achieved if $S_{int}[\phi,\Lambda]$ satisfies the renormalization group equation given in equation \eqref{ss} below, which is the Polchinski's ERG flow equation obtained in \cite{POLCHINSKI1984269}.
\\
\\
\noindent Alternatively in terms of functional derivatives acting on $S_{int}^{\Lambda_0}[\phi]$, the UV-regularized generating function above can also be written as follows \cite{golner2000exact,Ellwanger_1993,Ellwanger_1994},

\begin{align}
\label{apply1}
&e^{-W_{\Lambda}^{\Lambda_0}[J]}=\int D\phi e^{-\int_p\frac{1}{2}\phi(p)G^{-1}_B(\frac{p}{\Lambda})\phi(-p)+\int_pJ(p)\phi(-p)-S_{int}^{\Lambda_0}[\phi]}\\\nonumber
&=e^{-\frac{1}{2}\int_pJ(p)G_B(\frac{p}{\Lambda})J(-p)}e^{D_{\Lambda}^{\Lambda_0}}e^{-S_{int}^{\Lambda_0}[\phi]}\bigg|_{\phi=G_{B}J(p)},
\end{align}

\noindent where

\begin{align}
D_{\Lambda}^{\Lambda_0}=&\frac{1}{2}\int_pG_B\bigg(\frac{p}{\Lambda}\bigg)\frac{\delta}{\delta\phi(p)}\frac{\delta}{\delta\phi(-p)}\\\nonumber
=&\frac{1}{2}\int\frac{d^dp}{(2\pi)^d}G_B\bigg(\frac{p}{\Lambda}\bigg)\frac{\delta}{\delta\phi(p)}\frac{\delta}{\delta\phi(-p)}.
\end{align}

\noindent From the functional integral above, an effective interaction $S_{int}[\phi,\Lambda]$ can now be defined as,

\begin{align}
\label{effect}
e^{-S_{int}[\phi,\Lambda]}=e^{\int_pG_B(\frac{p}{\Lambda})\frac{\delta}{\delta\phi(p)}\frac{\delta}{\delta\phi(-p)}}e^{-S_{int}^{\Lambda_0}[\phi]},
\end{align}

\noindent where $S_{int}[\phi,\Lambda]$ consists of all Feynman diagrams generated by the interactions present in $S_{int}^{\Lambda_0}[\phi]$ which are obtained in the bare theory of the action. However the internal propagators involve both the UV cutoff $\Lambda_0$ and the infrared cutoff $\Lambda$. Now, taking a differential with respect to the infrared cutoff scale $\Lambda$ in equation \eqref{effect} above as shown below,

\begin{align}
\frac{\partial S_{int}[\phi,\Lambda]}{\partial\Lambda}e^{-S_{int}[\phi,\Lambda]}=\frac{\partial}{\partial\Lambda}\bigg(e^{\int_pG_B(\frac{p}{\Lambda})\frac{\delta}{\delta\phi(p)}\frac{\delta}{\delta\phi(-p)}}e^{-S_{int}^{\Lambda_0}[\phi]}\bigg),
\end{align}

\noindent the following differential flow equation is obtained below,

\begin{widetext}
\begin{align}
\label{ss}
-\Lambda\frac{\partial S_{int}[\phi,\Lambda]}{\partial \Lambda}=\frac{1}{2}\int_pG_{B}^\prime\big(\frac{p}{\Lambda}\big)\bigg(\frac{\delta^2S_{int}[\phi,\Lambda]}{\delta\phi(p)\delta\phi(-p)}-\frac{\delta S_{int}[\phi,\Lambda]}{\delta\phi(p)}\frac{\delta S_{int}[\phi,\Lambda]}{\delta\phi(-p)}\bigg),
\end{align}
\end{widetext}

\noindent where $G_{B}^\prime\big(\frac{p}{\Lambda}\big)=\Lambda\frac{\partial G_{B}(\frac{p}{\Lambda})}{\partial\Lambda}$, and the equation obtained above is the Polchinski's ERG flow equation \cite{POLCHINSKI1984269}. Building from these results above, we will now add spinless Fermionic fields with propagator  $G_F\big(\frac{p}{\Lambda}\big)$ to the functional integral as follows,

\
\begin{align}
&e^{-W_{\Lambda}^{\Lambda_0}[J]}=\int D\psi D\bar{\psi}D\phi e^{-\frac{1}{2}\int_p\phi(p)G_B^{-1}(\frac{p}{\Lambda})\phi(-p)}\\\nonumber
&\times e^{-\int_p\bar{\psi}(p)G_F^{-1}(\frac{p}{\Lambda})\psi(-p)+\int_pJ(p)\phi(-p)+\int_p\bar{\eta}(p)\psi(-p)+\int_p\eta(p)\bar{\psi}(-p)}\\\nonumber
&\times e^{- S_{int}^{\Lambda_0}[\phi,\bar{\psi},\psi]}
\end{align}
\

\noindent where $S_{int}^{\Lambda_0}[\phi,\bar{\psi},\psi]$ is the bare interaction part of the theory for the Bosonic and Fermionic fields, for simplicity we denote the bare action as $S_{int}^{\Lambda_0}[\phi,\bar{\psi},\psi]=S_{int}[\phi,\bar{\psi},\psi]$. Applying equation \eqref{apply1} to the Bosonic fields on the functional integral above we obtain,

\begin{align}
e^{-W_{\Lambda}^{\Lambda_0}[J]}&=\int D\psi D\bar{\psi}e^{-\int_p\bar{\psi}(p)G_F^{-1}(\frac{p}{\Lambda})\psi(-p)+\int_p\bar{\eta}(p)\psi(-p)}\\\nonumber
&\times e^{\int_p\eta(p)\bar{\psi}(-p)-\frac{1}{2}\int_pJ(p)G_B(\frac{p}{\Lambda})J(-p)}\\\nonumber
&\times e^{\frac{1}{2}\int_pG_B(\frac{p}{\Lambda})\frac{\delta}{\delta\phi(p)}\frac{\delta}{\delta\phi(-p)}e^{-S_{int}[\phi,\bar{\psi},\psi]}}\bigg|_{\phi=G_B(\frac{p}{\Lambda})J(p)}.\\\nonumber
\end{align}

\noindent Apply the same equation \eqref{apply1} to the remaining Fermionic fields in the functional integral we obtain

\begin{align}
&e^{-W_{\Lambda}^{\Lambda_0}[J]}=e^{-\frac{1}{2}\int_pJ(p)G_B(\frac{p}{\Lambda})J(-p)}\\\nonumber
&\times\bigg[e^{-\int_p\bar{\eta}(p)G_F\eta(-p)}e^{-\int_pG_F\frac{\delta}{\delta\bar{\psi}(p)}\frac{\delta}{\delta\psi(-p)}}\\\nonumber
&\times e^{-\frac{1}{2}\int_pG_B(\frac{p}{\Lambda})\frac{\delta}{\delta\phi(p)}\frac{\delta}{\delta\phi(-p)}}e^{-S_{int}[\phi,\bar{\psi},\psi]}\bigg]\bigg|_{\phi=G_BJ,\bar{\psi}=G_F\eta,\psi=G_F\bar{\eta}}.
\end{align}

\noindent Now we take the effective interaction action $S_{int}[\phi,\bar{\psi},\psi,\Lambda]$ from the functional integral above as,

\begin{align}
&e^{-S_{int}[\phi,\bar{\psi},\psi,\Lambda]} = e^{-\frac{1}{2}\int_pJ(p)G_B(\frac{p}{\Lambda})J(-p)}e^{-\int_p\bar{\eta}(p)G_F(\frac{p}{\Lambda})\eta(-p)}\\\nonumber
&\times e^{-\int_pG_F(\frac{p}{\Lambda})\frac{\delta}{\delta\bar{\psi}(p)}\frac{\delta}{\delta\psi(-p)}}e^{-\frac{1}{2}\int_pG_B(\frac{p}{\Lambda})\frac{\delta}{\delta\phi(p)}\frac{\delta}{\delta\phi(-p)}}e^{-S_{int}[\phi,\bar{\psi},\psi]},
\end{align}

\noindent and the flow equation for the theory is obtained by taking the derivative with respect the scale $\Lambda$,

\begin{align}
\label{erge_bf}
\frac{\partial}{\partial\Lambda}e^{-S_{int}[\phi,\bar{\psi},\psi,\Lambda]}&=\frac{\partial}{\partial\Lambda}\big(e^{-\frac{1}{2}\int_pJ(p)G_B(\frac{p}{\Lambda})J(-p)}\\\nonumber
&\times e^{-\int_p\bar{\eta}(p)G_F(\frac{p}{\Lambda})\eta(-p)}e^{-\int_pG_F(\frac{p}{\Lambda})\frac{\delta}{\delta\bar{\psi}(p)}\frac{\delta}{\delta\psi(-p)}}\\\nonumber
&\times e^{-\frac{1}{2}\int_pG_B(\frac{p}{\Lambda})\frac{\delta}{\delta\phi(p)}\frac{\delta}{\delta\phi(-p)}}e^{-S_{int}[\phi,\bar{\psi},\psi]}\big),\\\nonumber
\end{align}

\noindent and in the end the results simplifies to the following equation,

\begin{widetext}
\begin{align}
\label{general_polchinski}
-\Lambda\frac{\partial S_{int}}{\partial\Lambda}=&\int_{p} G_F^\prime\big(\frac{p}{\Lambda}\big)\bigg(\frac{\delta S_{int}}{\delta\bar{\psi}(p)}\frac{\delta S_{int}}{\delta\psi(-p)}-\frac{\delta^2S_{int}}{\delta\bar{\psi}(p)\delta\psi(-p)}\bigg)+\frac{1}{2}\int_{p} G_B^\prime\big(\frac{p}{\Lambda}\big)\bigg(\frac{\delta S_{int}}{\delta\phi(p)}\frac{\delta S_{int}}{\delta\phi(-p)}-\frac{\delta^2S_{int}}{\delta\phi(p)\delta\phi(-p)}\bigg),
\end{align}
\end{widetext}

\noindent which is the Polchinski ERG flow equation for an interaction theory between the Bosonic and spinless Fermionic fields. The effective interaction $S_{int}$ sometimes referred to as the interaction function or vertex function will be further defined as follows,

\begin{widetext}
\begin{align}
\label{general_int}
S_{int}[\phi,\bar{\psi},\psi,\Lambda] =& \sum_{m,n}^\infty\int\prod_{i=1}^{m}\frac{d^dk_i}{(2\pi)^d}\int\prod_{j=1}^{n}\frac{d^dp_jd^dq_j}{(2\pi)^{2d}}\phi(k_1)\cdots\phi(k_m)\big(\bar{\psi}(p_1)\psi(q_1)\big)\cdots\big(\bar{\psi}(p_n)\psi(q_n)\big)\\\nonumber
&\times\delta^d(\sum_i^mk_i+\sum_i^np_i+\sum_i^nq_i)\mathcal{L}_{m,n}(k_1,\cdots,k_m;p_1,\cdots,p_n;q_1,\cdots,q_n,\Lambda).
\end{align}
\end{widetext}

\noindent The first step of Wilson's Renormalization Group program is now complete. The first step already performed above, involves integrating out the high momentum modes of the fields, which is known as coarse-graining or blocking, this generates an equivalent effective action with reduced physical cutoff scale. Now, this has to be followed by the second step which is rescaling of the theory's action parameters (variables), by rescaling the momenta back to the original physical cutoff scale including appropriate rescaling of fields in order for the resulting equations to be cutoff scale independent, which also ensures the equations (theory) to have fixed points solutions relating to scale-invariant, massless renormalized theory. In the first step, the fields $\phi(p)$ are integrated over the range $e^{-t}<|p|\leq 1$ such that the partition function is left invariant, and the second step involves rescaling, a change by a factor of $e^{-t}$ in all linear dimensions to restore the original scale $\Lambda$ of the system, i.e $p\rightarrow p^\prime = e^tp$, where the parameter $t$ is an infinitesimal parameter.

\noindent Adopting the approach used in \cite{Bagnuls_2001, Rosten_2012} to apply rescaling on Polchinski's ERG flow equation shown earlier in equation \eqref{general_polchinski}, results in Polchinski's original equation having some additional terms coming from applying rescaling on the original equation parameters, below we demonstrate how the additional terms emanate. First, for simplicity we drop the Fermionic fields from equation \eqref{general_int}, then the action  $S_{int}$ becomes,

\begin{align}
\label{no_fermions}
S[\phi,\Lambda]=&\sum_{m}\int\prod_{i=1}^m\frac{d^dp_i}{(2\pi)^d}\mathcal{L}_{m,0}(p_1,\cdots ,p_m,\Lambda)\\\nonumber
&\times \phi(p_1)\cdots\phi(p_m)\delta^d(p_1+\cdots +p_m).
\end{align}

\noindent Under rescaling the momentum and Bosonic fields will change by a factor $s=e^{-t}$ in all linear dimensions, and they respectively transform as,

\begin{align}
&p=(1+t)p\\\nonumber
&\phi(s^{-1}p^\prime)=(1+t)^{d-d_\phi}\phi(p),
\end{align}

\noindent and applying this form of rescaling on Polchinski's ERG flow equation \eqref{general_polchinski} for a case of single self-interacting Bosonic fields leads to the full rescaled Polchinski's ERG flow equation,

\begin{widetext}
\begin{align}
\label{re_bos}
&\big(\partial_{t}+d_\phi \bigtriangleup_{\phi}+\bigtriangleup_{\partial_\phi}-d\big)S[\phi,t] =
\frac{1}{2}\int_{p} G_B^\prime\big(\frac{p}{\Lambda}\big)\bigg(\frac{\delta S[\phi,t]}{\delta\phi(p)}\frac{\delta S[\phi,t]}{\delta\phi(-p)}-\frac{\delta^2S[\phi,t]}{\delta\phi(p)\delta\phi(-p)}\bigg),
\end{align}
\end{widetext}

\noindent where 
$\partial_t=\Lambda\frac{\partial}{\partial\Lambda}$ is the time derivative,
\begin{align}
&\Delta_{\partial_{\phi}}=d+\int_p\phi_pp\cdot \partial_{p}\frac{\delta}{\delta\phi_p},\\\nonumber
&\Delta_{\phi}=\int_p\phi_p\frac{\delta}{\delta\phi_p},\\\nonumber 
&d_{\phi}=\frac{1}{2}(d-2+\eta_{\phi})
\end{align}

\noindent and $\eta_{\phi}$ is the anomalous dimension of the scalar field $\phi(p)$. The details regarding how the results in equation \eqref{re_bos} were obtained are provided in the Appendix. Treating the Fermionic fields in similar fashion, in the case of self-interacting spinless Fermionic fields, now $S_{int}$ will be $S[\psi,\bar{\psi}, \Lambda]$, and the final derivation results for the Polchinski's ERG flow equation is,

\begin{widetext}
\begin{align}
\big(\partial_t-d+d_{\psi\bar{\psi}}\Delta_{\psi\bar{\psi}}+\Delta_{\partial_{\psi\bar{\psi}}}\big)S[\psi,\bar{\psi},t]=\int_{p} G_F^\prime\big(\frac{p}{\Lambda}\big)\bigg(\frac{\delta S[\psi,\bar{\psi},t]}{\delta\psi(p)}\frac{\delta S[\psi,\bar{\psi},t]}{\delta\bar{\psi}(-p)}-\frac{\delta^2S[\psi,\bar{\psi},t]}{\delta\psi(p)\delta\bar{\psi}(-p)}\bigg).
\end{align}

\end{widetext}

\noindent where
\begin{align}
&\Delta_{\partial_{\psi\bar{\psi}}}=d+\int\frac{d^dp}{(2\pi)^d}\psi(p)p\cdot\frac{\partial}{\partial p}\frac{\delta}{\delta\psi(-p)}\\\nonumber
&\quad+\int\frac{d^dp}{(2\pi)^d}\bar{\psi}(p)p^\cdot\frac{\partial}{\partial p}\frac{\delta}{\delta\bar{\psi}(-p)}
\end{align}
\begin{align}
&\Delta_{\psi\bar{\psi}}=\int_p\psi(p)\frac{\delta}{\delta\psi(-p)}+\int_p\bar{\psi}(p)\frac{\delta}{\delta\bar{\psi}(-p)}\\\nonumber
&d_{\psi\bar{\psi}}=\frac{1}{2}(d-1+\eta_{\psi})
\end{align}
and $\eta_{\psi}$ is the anomalous dimensions of the Fermionic fields. Now, the general rescaled Polchinski's ERG flow equation for the interactions between the single Bosonic and spinless Fermionic fields together will be,

\begin{widetext}
\begin{align}
\label{integro}
\big(\partial_t+d_{\psi\bar{\psi}}\Delta_{\psi\bar{\psi}}+\Delta_{\partial_{\psi\bar{\psi}}}+&d_\phi \bigtriangleup_{\phi}+\bigtriangleup_{\partial_\phi}-d\big)S[\phi, \psi,\bar{\psi},t]=\int_{p} G_F^\prime\big(\frac{p}{\Lambda}\big)\bigg(\frac{\delta S[\phi,\psi,\bar{\psi},t]}{\delta\psi(p)}\frac{\delta S[\phi,\psi,\bar{\psi},t]}{\delta\bar{\psi}(-p)}-\frac{\delta^2S[\phi,\psi,\bar{\psi},t]}{\delta\psi(p)\delta\bar{\psi}(-p)}\bigg)\\\nonumber
&+\frac{1}{2}\int_{p} G_B^\prime\big(\frac{p}{\Lambda}\big)\bigg(\frac{\delta S[\phi,\psi,\bar{\psi},t]}{\delta\phi(p)}\frac{\delta S[\phi,\psi,\bar{\psi},t]}{\delta\phi(-p)}-\frac{\delta^2S[\phi,\psi,\bar{\psi},t]}{\delta\phi(p)\delta\phi(-p)}\bigg).
\end{align}
\end{widetext}

\noindent This is the final derivation we were after in this section. We have derived the Polchinski's ERG flow equation for the single Bosonic and spinless Fermionic fields. We chose spinless Fermionic fields since their computations are much easier to handle here, and the results can be generalized to Fermionic fields with spins.

\section{Bosonic Field Local Potential Approximation.}
\noindent The Polchinski's ERG flow equation derived in the previous section in equation \eqref{integro} is an integro-differential equation, in general this type of differential equation is challenging to solve, especially at non-perturbative level. In Renormalization Group one can handle such type of differential equations by making use of perturbative expansion either in powers of small coupling parameter epsilon ($\epsilon$) expansion or $\frac{1}{N}$ large $N$ parameter expansion. At non-perturbative level and in the absence of expansion parameters, one way to tackle this differential equation is to do some approximation or truncation by reducing the number of degrees of freedom. Performing derivative or momentum expansion of the effective action is one possible approximation that can be undertaken, therefore here we will take the effective action $S_{int}$ from equation \eqref{general_int} with only Bosonic fields,

\begin{align}
S[\phi,\Lambda]=&\sum_{m}\int\prod_{i=1}^m\frac{d^dk_i}{(2\pi)^d}\mathcal{L}_{m,0}(k_1,\cdots ,k_m,\Lambda)\\\nonumber
&\times \phi(k_1)\cdots\phi(k_m)(2\pi)^d\delta^d(k_1+\cdots +k_m),
\end{align}

\noindent and expand the vertex or coupling function $\mathcal{L}_{m,0}$ in terms of $k^2$ momentum, which is a derivative expansion in position space,

\begin{align}
\mathcal{L}_{m,0}(k_1,&\cdots,k_m,\Lambda)=\mathcal{L}_{m,0}(0,\cdots,0,\Lambda)\\\nonumber
&+\frac{1}{m(m-1)}(k_1^2+\cdots +k_m^2)\mathcal{L}_{m-2,0}(0,\cdots,0,\Lambda)\\\nonumber
&+\mathcal{O}(k^4)+\cdots.
\end{align}

\noindent It then follows that the effective action $S[\phi,\Lambda]$ will be an expansion in $k^2$ momentum,

\begin{widetext}
\begin{align}
S[\phi,\Lambda] =& \sum_{m}\int\prod_{i=1}^m\frac{d^dk_i}{(2\pi)^d}\bigg( \mathcal{L}_{m,0}+\frac{k_1^2+\cdots+k_m^2}{m(m-1)}\mathcal{L}_{m-2,0}+\mathcal{O}(k^4)+\cdots \bigg)\phi(k_1)\cdots\phi(k_m)(2\pi)^d\delta^d(k_1+\cdots +k_m)\\\nonumber
=&\int\prod_{i=1}^m\frac{d^dk_i}{(2\pi)^d}\bigg(V(\phi,\Lambda)+W(\phi,\Lambda)+\mathcal{O}(k^4)\bigg)
\end{align}

\noindent where
\begin{align}
V(\phi,\Lambda)=\mathcal{L}_{m,0}(0,\cdots,0,\Lambda)\phi(k_1)\cdots\phi(k_m)(2\pi)^d\delta^d(k_1+\cdots +k_m)
\end{align}
\end{widetext}
is a local term which contains no derivatives, this term can be viewed as a local potential term, while the $W(\phi,\Lambda)$ term and other higher order expansion terms are non-local, these terms have derivatives acting on them in position space. To project out only the local potential term from the action $S[\phi,\Lambda]$ we can make use of the Hasenfratz and Hasenfratz projection operator $P(z)$ \cite{Hasenfratz:1985dm}. When this projection operator is applied on the action $S[\phi,\Lambda]$ it only selects the local potential term as follows,

\begin{align}
\label{r11}
P(z)S[\phi,\Lambda]=&e^{z\frac{\partial}{\partial\phi(0)}}S[\phi,\Lambda]\bigg|_{\phi=0}\\\nonumber
=&\sum_m\mathcal{L}_{m,0}(0,0,\cdots ,0,\Lambda)z^m\delta^d(0)\\\nonumber
=&V(z,\Lambda)\delta^d(0),
\end{align}

\noindent where $V(z,\Lambda)$ is a local potential function. Then the action of the operator $P(z)$ on the term $\frac{\delta S[\phi,\Lambda]}{\delta \phi}$ yields the results,

\begin{align}
\label{r2}
\int_pP(z)\frac{\delta S[\phi,\Lambda]}{\delta\phi(p)}\bigg|_{\phi=0}=&\int_pe^{z\frac{\partial}{\partial\phi(0)}}\frac{S[\phi,\Lambda]}{\delta\phi(p)}\bigg|_{\phi=0}\\\nonumber
=&V^\prime(z,\Lambda)\delta^d(p),
\end{align}

\noindent where  $V^\prime(z,\Lambda)$ is the derivative of the local potential function $V(z,\Lambda)$ with respect to $z$. Therefore, when the operator $P(z)$ acts on the entire rescaled Polchinski's ERG flow equation as follows,

\begin{widetext}
\begin{align}
P(z)\big(\partial_t+d_{\phi}\triangle_{\phi}+\triangle_{\partial}-d\big)S[\phi,t]=P(z)\bigg(\int_{p} G_B^\prime\big(\frac{p}{\Lambda}\big)\bigg(\frac{\delta S[\phi,t]}{\delta\bar{\phi}(p)}\frac{\delta S[\phi,t]}{\delta\phi(-p)}-\frac{\delta^2S[\phi,t]}{\delta\bar{\phi}(p)\delta\phi(-p)}\bigg)\bigg),
\end{align}

\noindent it then leads to the following results,

\begin{align}
\bigg(\partial_t+d_{\phi}\triangle_{\phi}+\triangle_{\partial}-d\bigg)V(z,t)&= G_B^\prime(0)V^{\prime 2}(z,t)-V^{\prime\prime}(z,t)\int_p G_B^\prime\big(\frac{p}{\Lambda}\big),
\end{align}
\end{widetext}

\noindent where now $d_{\phi}=\frac{d-2}{2}$ and the anomalous dimension $\eta_\phi =0$ in the LPA limit since there is no momentum dependence in the zeroth order of the derivative expansion. Further simplification leads to the following differential equation,

\begin{align}
\partial_tV(z,t)=&\int_p G_B^\prime\big(\frac{p}{\Lambda}\big)V^{\prime\prime}(z,t)- G_B^\prime(0)V^{\prime 2}(z,t)\\\nonumber
&-d_{\phi}zV^{\prime}(z,t)+dV(z,t),\\\noindent
\end{align}
\

\noindent and writing $V(z,t)$ and $z$ as follows,

\begin{align}
V(z,t)\rightarrow& \frac{V(z,t)\int_pG_B^\prime(\frac{p}{\Lambda})}{G_B^\prime(0)},\\\nonumber
z\rightarrow& z\sqrt{\int_pG_B^\prime\big(\frac{p}{\Lambda}\big)},\\\nonumber
\end{align}

\noindent then the differential equation above becomes,

\begin{align}
\label{x1}
\partial_tV(z,t)=V^{\prime\prime}(z,t)-V^{\prime 2}(z,t)-d_{\phi}zV^{\prime}(z,t)+dV(z,t).
\end{align}

\noindent To find the fixed points of the equation above where $V(z,t)=V_*(z,t)$ we have to calculate $\partial_tV(z,t)=0$. We begin with a simple approach, first we opt to remove the rescaling terms from the Polchinski's ERG flow equation. We will drop the terms $-d_{\phi}zV^{\prime}(z,t)+dV(z,t)$ from equation \eqref{x1}, these terms emanate from applying rescaling on the Polchinski's ERG flow equation. Dropping of rescaled terms results in the following Polchinski's ERG flow equation,

\begin{align}
\partial_tV(z,t)= V^{\prime\prime}(z,t)-V^{\prime 2}(z,t)\\\noindent
\partial_t e^{-V(z,t)}=\frac{\partial^2}{\partial z\partial z}e^{-V(z,t)}.
\end{align}

\noindent Now let $U(z,t)=e^{-V(z,t)}$, then $V(z,t)=-\ln(U(z,t))$, and the differential equation above will now be written as follows
\begin{align}
\label{homog}
\partial_tU(z,t)=\frac{\partial^2}{\partial z\partial z}U(z,t),
\end{align}

\noindent which resembles a homogeneous differential equation for the heat equation, and the solution is of the form,
\begin{align}
\label{homo_sol}
&U(z,t)=\frac{1}{\sqrt{4\pi\lambda t}}e^{-\frac{z^2}{4\lambda t}},\\\nonumber
&V(z,t)=-\ln(U(z,t)).
\end{align}

\noindent The fixed points of this equation are obtained from computing $\partial_tU_*(z,t)=0$ or $\partial_tV_*(z,t)=0$, which gives the following solution,

\begin{align}
\label{homo_solution}
&\partial_tV_*(z,t)=0\\\nonumber
&\Rightarrow t=\frac{z^2}{2\lambda}\qquad\qquad\qquad\qquad\\\nonumber
&\Rightarrow U_*(z,t)=\frac{1}{z\sqrt{2\pi}}e^{-\frac{1}{2}}\\\nonumber
&\Rightarrow V_*(z,t) = \ln\big(z\sqrt{2\pi}\big)+\frac{1}{2}.
\end{align}

\noindent The solution indicates that at the fixed point we have a logarithmic potential or interaction (vertex) function $V_*(z,t)=\ln(z\sqrt{2\pi})+\frac{1}{2}$. Which could give an indication that the fixed point is governed by the logarithmic CFT (Conformal Field Theory), and when $t\rightarrow 0$ we see that $U(z,t)\rightarrow 0$ and $V(z,t)\rightarrow \infty$. Now, using the full rescaled Polchinski's ERG flow equation in equation \eqref{x1},

\begin{align}
\label{rescaled_pol}
\partial_tV(z,t)=V^{\prime\prime}(z,t)-V^{\prime 2}(z,t)-d_{\phi}zV^{\prime}(z,t)+dV(z,t),
\end{align}
\noindent and perturbing around the fixed point solution to find non trivial fixed points solution at order $\mathcal{O}(g_b)$ where the perturbed solution $V_p(z,t)$ is given as,

\begin{align}
\label{pert_lin}
V_p(z,t)=&V(z,t)+g_b\delta V(z,t)\\\nonumber
=&V(z,t)+g_b\sum_nf_n(z)e^{\lambda_n t}
\end{align}
\noindent where $V(z,t)$ is the homogeneous solution obtained in equation \eqref{homo_sol}, and $g_b\ll 1$ is a small interaction constant or term. Plugging $V_p(z,t)$ into the full rescaled Polchinski's differential equation obtained within the LPA limit in equation \eqref{x1},
\begin{widetext}
\begin{align}
\label{pert_lin2}
\partial_tV_p(z,t)= V^{\prime\prime}_p(z,t)-V^{\prime 2}_p(z,t)-d_{\phi}zV^{\prime}_p(z,t)+dV_p(z,t),
\end{align}

\noindent at order $\mathcal{O}(g_b)$ the differential equation below is obtained,\\

\begin{align}
\label{epsilon-ex2}
e^{\lambda t}\frac{\partial^2f(z)}{\partial z\partial z}=e^{\lambda t}\bigg((\lambda-d)f(z)+d_{\phi}z\frac{\partial f(z)}{\partial z}+2\frac{\partial V(z,t)}{\partial z}\frac{\partial f(z)}{\partial z}\bigg),
\end{align}
\end{widetext}

\noindent expanding the last term on the RHS as shown below,

\begin{align}
\label{boson_expansion}
e^{\lambda t}\frac{\partial V(z,t)}{\partial z}\frac{\partial f(z)}{\partial z}=
\frac{z}{2}\bigg(\frac{1}{\lambda t}+1+\frac{\lambda t}{2!}+\cdots\bigg)\frac{\partial f(z)}{\partial z},
\end{align}
\noindent and then collecting the zeroth order ($\mathcal{O}(t^0)$) terms in expansion parameter $t$ from the differential equation \eqref{epsilon-ex2} we get the following differential equation below,

\begin{align}
\frac{\partial^2f(z)}{\partial z\partial z}=&(\lambda-d)f(z)+d_{\phi}z\frac{\partial f(z)}{\partial z}+2\frac{\lambda}{\lambda} \frac{z}{2}\frac{\partial f(z)}{\partial z}\\\nonumber
=&(\lambda-d)f(z)+\big(d_{\phi}+1\big)z\frac{\partial f(z)}{\partial z}\\\nonumber
=&(\lambda-d)f(z)+c_{\phi\lambda}z\frac{\partial f(z)}{\partial z},
\end{align}

\noindent where $c_{\phi\lambda}=d_{\phi}+1$. Writing the function $f(z)$ as $f(z)\rightarrow \frac{2}{c_{\phi\lambda}}f(\sqrt{\frac{c_{\phi\lambda}}{2}} z)$ \cite{Comellas_1998,cmp/1104159469,Bagnuls_2001,Hasenfratz:1985dm}, then the differential equation above simplifies to,
\begin{align}
\frac{\partial^2f(\tilde{z})}{\partial \tilde{z}\partial \tilde{z}}= 2\frac{\lambda-d}{c_{\phi\lambda}}f(\tilde{z})+2\tilde{z}\frac{\partial f(\tilde{z})}{\partial \tilde{z}},
\end{align}

\noindent where $\tilde{z}=\sqrt{\frac{c_{\phi\lambda}}{2}} z$. This is a Hermite differential equation, which is analogous to the Hermite differential equation also obtained in \cite{Comellas_1998,cmp/1104159469,Hasenfratz:1985dm,Bagnuls_2001}, and the Hermite polynomial which is the solution to this differential equation is given as follows,

\begin{align}
\label{boson_polyn}
f(z)=f_k^{(0)}(\tilde{z})=\sum_{r=0}^{\frac{k}{2}}\frac{(-1)^rk!}{r!(k-2r)!}(2\tilde{z})^{k-2r},
\end{align}

\noindent with the Hermite polynomial eigenvalues $\lambda=\lambda_k^{(0)}$ given as,

\begin{align}
\label{eq_zero}
&2\frac{d-\lambda_k^{(0)}}{c_{\phi\lambda}}=2k-1,\quad k=0,1,2,3,\cdots\\\nonumber
&\Rightarrow \lambda_k^{(0)}= -\frac{(2k-1)d}{4}+d,
\end{align}

\noindent where the superscript $(0)$ indicates that these are zeroth order expansion in parameter $t$ eigenvalues and eigenfunctions. The Hermite polynomials are the eigenfunctions or eigenoperators of the theory at this limit. Expanding equation \eqref{epsilon-ex2} again to first order in parameter $t$, the Hermite differential equation reads as follows,

\begin{align}
\frac{\partial^2f(z)}{\partial z\partial z}=&(\lambda-d)f(z)+d_{\phi}z\frac{\partial f(z)}{\partial z}+2\frac{\lambda}{2\lambda} \frac{z}{2}\frac{\partial f(z)}{\partial z}\\\nonumber
=& 2\frac{\lambda-d}{c_{\phi\lambda}}f(z)+2z\frac{\partial f(z)}{\partial z},
\end{align}

\noindent where this time $f(z)\rightarrow \frac{2}{c_{\phi\lambda}}f\bigg(\sqrt{\frac{c_{\phi\lambda}}{2}} z\bigg)$ and here $c_{\phi\lambda}=d_{\phi}+\frac{1}{2}$. The corresponding Hermite polynomial at this first order in expansion parameter $t$ will be the same as the previous one given in equation \eqref{boson_polyn},

\begin{align}
f(z)=f_k^{(1)}(z)=\sum_{r=0}^{\frac{k}{2}}\frac{(-1)^rk!}{r!(k-2r)!}(2z)^{k-2r},
\end{align}

\noindent with the eigenvalues $\lambda=\lambda_k^{(1)}$ for the Hermite polynomial at this order given as,

\begin{align}
\label{eq_first}
\lambda_k^{(1)}=-\frac{(2k-1)(d-1)}{4}+d, \quad k=0,1,2,3,\cdots.
\end{align}

\noindent Then at the $n^{th}$ order in expansion parameter $t$ we get the following Hermite differential,

\begin{align}
\frac{\partial^2f(z)}{\partial z\partial z}=& 2\frac{\lambda-d}{c_{\phi\lambda}}f(z)+2z\frac{\partial f(z)}{\partial z},
\end{align}

\noindent where now $c_{\phi\lambda}=d_{\phi}+\frac{1}{(n+1)}$, and the Hermite polynomial is the same as before, with the eigenvalues $\lambda=\lambda_k^{(n)}$ at $n^{th}$ order given as,

\begin{align}
\label{eq_infty}
\lambda_k^{(n)}=-\frac{(2k-1)(d-2+\frac{2}{n+1})}{4}+d.
\end{align}

\noindent Then at $n\rightarrow \infty$,

\begin{align}
\lambda_k^{(\infty)}=-\frac{(2k-1)(d-2)}{4}+d.
\end{align}

\noindent Take note that at order $\frac{1}{t}$ expansion we have set the function $f^{(-1)}(z)=constant$, since when expanding \eqref{epsilon-ex2} we find that,

\begin{align}
&0=\frac{z}{\lambda t}\frac{\partial f^{(-1)}(z)}{\partial z}\\\nonumber
&\Rightarrow f^{(-1)}(z)=constant.
\end{align}

\noindent Putting together all the expanded terms above originating from equation \eqref{epsilon-ex2} we arrive at the solution,

\begin{widetext}
\begin{align}
V_p(z,t)=&V(z,t)+g_b e^{\lambda t}f(z)\\\nonumber
=&V(z,t)+g_b\bigg[\frac{1}{\lambda t}f^{(-1)}(z)+\lambda_k^{(0)}f_k^{(0)}(z)+t\lambda_k^{(1)}f_k^{(1)}(z)
+\frac{t^2}{2!}\lambda_k^{(2)}f_k^{(2)}(z)+\cdots+\frac{t^n}{n!}\lambda_k^{(n)}f_k^{(n)}(z)\bigg]\\\nonumber
=&V(z,t)+\frac{g_b}{\lambda t}f^{(-1)}_k(z)+\sum_{n=0}\frac{\lambda_k^{(n)}t^{n}}{n!}f_k^{(n)}(z).
\end{align}

\noindent Then at the logarithmic fixed point at $t=\frac{z^2}{2\lambda}$, the solution above becomes,

\begin{align}
\label{b_results}
V_p(z,t) =&\ln\big(z\sqrt{\pi}\big)+\frac{1}{2}+g_b\bigg(\frac{2}{z^2}f^{(-1)}(z)+\lambda_k^{(0)}f_k^{(0)}(z)+\frac{z^2}{2\lambda}\lambda^{(1)}_kf_k^{(1)}(z)+\cdots+\frac{z^{2n}}{(2\lambda)^nn!}\lambda_k^{(n)}f_k^{(n)}(z)\bigg),\\\nonumber
=&\ln\big(z\sqrt{\pi}\big)+\frac{1}{2}+2\frac{g_b}{z^2}f^{(-1)}(z)+\sum_{n=0}\frac{\lambda_k^{(n)}}{n!}\frac{z^{2n}}{(2\lambda)^n}f_k^{(n)}(z).
\end{align}
\end{widetext}

\noindent The results obtained here show that perturbing the interaction or vertex function in the vicinity of a logarithmic fixed point where we first linearize around the homogeneous solution as shown in equation \eqref{pert_lin}, plugging this interaction function into the full rescaled Polchinski's ERG flow equation as shown in equation \eqref{pert_lin2}, the solution leads to an infinite set tower of Hermite differential equation in expansion parameter $t$. The solution to these Hermite differential equations is a Hermite polynomial, but the Hermite eigenvalues for each Hermite differential equation are different at each order in expansion parameter $t$ as demonstrated in equations \eqref{eq_zero}, \eqref{eq_first} and \eqref{eq_infty}.\\
\\
\noindent Now that the fixed points for the Bosonic system have been determined, the natural step that follows should be the computation of the systems critical exponents. Here we will calculate the critical exponents $\nu$ and $w$, where $\nu$ is the correlation length exponent and is obtained from the highest positive eigenvalue ($\nu=\frac{1}{\lambda}$), while $w$ is the first correction-to-scaling and is obtained from the less negative eigenvalue. In instances where there are an infinity many negative eigenvalues, then $w_i$ will be $i^{th}$ correction-to-scaling whereby $w_1$ will denote the first correction-to-scaling, $w_2$ denoting the second correction-to-scaling etc. Here we are also going to calculate the critical points at each order in expansion parameter $t$. Starting with the zero order expansion in parameter $t$, at $d=4$, the eigenvalues for the eigenfunctions will be,

\begin{align}
\lambda_k^{(0)}=&-\frac{(2k-1)d}{4}+4\\\nonumber
=&-(2k-1)+4,\qquad k=0,1,2,\cdots,
\end{align}

\noindent and the non-negative eigenvalues are

\begin{align}
\lambda_0^{(0)}=5,\quad \lambda_1^{(0)}=3,\quad \lambda_2^{(0)}=1.
\end{align}

\noindent The corresponding eigenfunctions for these non-negative eigenvalues respectively will be,

\begin{align}
f_0^{(0)}=&1,\\\nonumber
f_1^{(0)}=&2z,\\\nonumber
f_2^{(0)}=&4z^2-2.
\end{align}

\noindent Now, the first two negative eigenvalues are $\lambda_3=-1$, $\lambda_4=-3$ and the corresponding eigenfunctions respectively are,

\begin{align}
\label{eigop}
f_3^{(0)}=&8z^3-12z,\\\nonumber
f_4^{(0)}=&16z^4-48z^2+12.
\end{align}

\noindent The first negative eigenvalues closest to zero are used to calculate the critical exponent $w_1$ and $w_2$, and we find,

\begin{align}
w_1=1,\\\nonumber
w_2=3.
\end{align}

\noindent The largest positive eigenvalue $\lambda_0^{(0)}=5$ is used to calculate the critical exponent $\nu$,
and we find,
\begin{align}
\nu=0.2.
\end{align}

\noindent Moving to the first order expansion in parameter $t$, the eigenvalue formula for the first order eigenfunction at $d=4$ is,
\begin{align}
\lambda_k^{(1)}=&-\frac{(2k-1)(d-1)}{4}+d,\\\nonumber
=&-(2k-1)\frac{3}{4}+4.
\end{align}

\noindent The non-negative eigenvalues are, $\lambda_0^{(1)}=\frac{19}{4}$, $\lambda_1^{(1)}=\frac{13}{4}$, $\lambda_2^{(1)}=\frac{5}{4}$ and their corresponding eigenfunctions respectively are the same as in equation \eqref{eigop},

\begin{align}
f_0^{(1)}=&1,\\\nonumber
f_1^{(2)}=&2z,\\\nonumber
f_2^{(3)}=&4z^2-2.
\end{align}

\noindent The first two negative eigenvalues are $\lambda_3^{(1)}=-\frac{1}{4}$ and $\lambda_4^{(1)}=-\frac{5}{4}$. At this order the critical exponents will be as follows,

\begin{align}
&\nu=0.21,\\\nonumber
&w_1=0.25,\\\nonumber
&w_2=1.25.
\end{align}

\noindent At the second order in expansion parameter $t$ we find the critical exponents to be,

\begin{align}
\nu=0.21,\quad w_1=0.67,\quad w_2=4.67.
\end{align}

\noindent At this point the process should now be straight forward pertaining to how the critical exponents can be calculated at each order in expansion parameter $t$, therefore we will stop the computations at this order. Due to the infinite set of Hermite differentials equations obtained earlier in equation \eqref{epsilon-ex2} where they had the same Hermite polynomial as the solution but with different eigenvalues at each order in expansion parameter $t$, this also naturally forces the critical exponents to have different values at each order in expansion parameter $t$.

\section{Fermion fields homogeneous and hermite differential equations}
\noindent For the Fermions we follow similar steps undertaken in the single Bosonic field computations from the previous section. We start by first writing down an operator that will extract and return a local potential term, similar to the one given in Bosonic fields in equation \eqref{r11}. For the Fermions we define the operator,

\begin{align}
\label{fermion_op}
P(z,\bar{z})X[\psi,\bar{\psi}]=e^{z\frac{\partial}{\partial\psi(0)}+\bar{z}\frac{\partial}{\partial\bar{\psi}(0)}}X[\psi,\bar{\psi}]\bigg|_{\psi,\bar{\psi}=0},
\end{align}

\noindent where $X[\psi,\bar{\psi}]$ is the test functional operator containing Fermionic fields, and the variables $z$ and $\bar{z}$ are Grassman variables. Dropping the Bosonic fields from the effective action term $S_{int}$ in equation \eqref{general_int}, $S_{int}$ will now be,

\begin{widetext}
\begin{align}
S[\psi,\bar{\psi},\Lambda] = &\sum_{n=1}\int\prod_{j=1}^n\frac{d^dp_jd^dq_j}{(2\pi)^d}\mathcal{L}_{0,n}(p_1,\cdots,p_n;q_1,\cdots,q_n,\Lambda)
\times\bar{\psi}(p_1)\psi(q_1)\cdots \bar{\psi}(p_n)\psi(q_n)\delta^d(\sum_i^np_i+\sum_i^nq_i),
\end{align}

\noindent and the full rescaled Polschinski's ERG flow equation for the Fermionic fields is,

\begin{align}
\bigg(\partial_t+d_{\psi\bar{\psi}}\triangle_{\psi\bar{\psi}}+\triangle_{\partial\psi\bar{\psi}}-d\bigg)S[\psi,\bar{\psi},t]=\int_p G_F^\prime\big(\frac{p}{\Lambda}\big)\bigg(\frac{\delta S[\psi,\bar{\psi},t]}{\delta\bar{\psi}(p)}\frac{\delta S[\psi,\bar{\psi},t]}{\delta\psi(-p)}-\frac{\delta^2S[\psi,\bar{\psi},t]}{\delta\bar{\psi}(p)\delta\psi(-p)}\bigg),
\end{align}
\end{widetext}

\noindent where 
\begin{align}
&\Delta_{\psi\bar{\psi}}=\int_p\psi(p)\frac{\delta}{\delta\psi(-p)}+\int_p\bar{\psi}(p)\frac{\delta}{\delta\bar{\psi}(-p)}\\\nonumber &\Delta_{\partial_{\psi\bar{\psi}}}=d+\int\frac{d^dp}{(2\pi)^d}\psi(p)p\cdot\frac{\partial}{\partial p}\frac{\delta}{\delta\psi(-p)}\\\nonumber
&\qquad\qquad+\int\frac{d^dp}{(2\pi)^d}\bar{\psi}(p)p^\cdot\frac{\partial}{\partial p}\frac{\delta}{\delta\bar{\psi}(-p)}
\end{align}
Acting with the operator $P(z,\bar{z})$ given in equation \eqref{fermion_op} on the differential equation above, so that the operator $P(z,\bar{z})$ extracts and returns only the local potential terms in $S[\psi,\bar{\psi},t]$ while at the same time the fields $\psi$ and $\bar{\psi}$ are replaced by the Grassmann variables $z$ and $\bar{z}$ respectively as shown below,

\begin{widetext}
\begin{align}
P(z,\bar{z})\bigg(\partial_t+d_{\psi\bar{\psi}}\triangle_{\psi\bar{\psi}}+\triangle_{\partial\psi\bar{\psi}}-d\bigg)S[\psi,\bar{\psi},t]=&P(z,\bar{z})\int_p G_F^\prime\big(\frac{p}{\Lambda}\big)\bigg(\frac{\delta S[\psi,\bar{\psi},t]}{\delta\bar{\psi}(p)}\frac{\delta S[\psi,\bar{\psi},t]}{\delta\psi(-p)}-\frac{\delta^2S[\psi,\bar{\psi},t]}{\delta\bar{\psi}(p)\delta\psi(-p)}\bigg)\\\nonumber
\partial_tV(z,\bar{z},t)+d_{\psi\bar{\psi}}\bigg(z\frac{\partial V(z,\bar{z},t)}{\partial z}+\bar{z}\frac{\partial V(z,\bar{z},t)}{\partial \bar{z}}\bigg)&-dV(z,\bar{z},t)=\bigg( G_F^\prime(0)\frac{\partial V(z,\bar{z},t)}{\partial z}\frac{\partial V(z,\bar{z},t)}{\partial\bar{z}}-\frac{\partial^2V(z,\bar{z},t)}{\partial z\partial\bar{z}}\int_p G_F^\prime\big(\frac{p}{\Lambda}\big)\bigg),
\end{align}
\end{widetext}

\noindent where $V(z,\bar{z},t)=\sum_n\mathcal{L}_{0,n}(0,\cdots,0;t)(z\bar{z})^n$, and for generality we are less focused on the fact that the higher degree terms of $V(z,\bar{z},t)$ will be nilpotent. Making similar transformations as in the Bosonic fields case where,

\begin{align}
V(z,\bar{z},t)\rightarrow& \frac{V(z,\bar{z},t)\int_pG_F^\prime(\frac{p}{\Lambda})}{G_F^\prime(0)}\\\nonumber
z\rightarrow& z\sqrt{\int_pG_F^\prime\big(\frac{p}{\Lambda}\big)}\\\nonumber
\bar{z}\rightarrow&\bar{z}\sqrt{\int_pG_F^\prime\big(\frac{p}{\Lambda}\big)},
\end{align}

\noindent and then dropping the terms emanating from rescaling the Polchinski's ERG flow equation simplifies to,

\begin{align}
\label{fermion_heat}
\partial_tV(z,\bar{z},t)=\frac{\partial^2V(z,\bar{z},t)}{\partial z\partial\bar{z}}-\frac{\partial V(z,\bar{z},t)}{\partial z}\frac{\partial V(z,\bar{z},t)}{\partial\bar{z}}.
\end{align}

\noindent Let $U(z,\bar{z},t)=e^{-V(z,\bar{z},t)}$, then equation \eqref{fermion_heat} becomes,

\begin{align}
\label{fermion_results_intro2}
\partial_t U(z,\bar{z},t)=\frac{\partial^2U(z,\bar{z},t)}{\partial z\partial\bar{z}},
\end{align}
\noindent which is a Grassmann variable analogue of a homogeneous differential equation obtained earlier in equation \eqref{homog}, and the solution is of the form,

\begin{align}
&U(z,\bar{z},t)=ce^{\frac{z\bar{z}}{\lambda t}+\ln(\lambda t)}\\\nonumber
\Rightarrow &V(z,\bar{z},t)=\frac{z\bar{z}}{\lambda t}+\ln(\lambda t)+\ln(c).
\end{align}

\noindent where $c$ is a constant, for simplicity we set it to a unit, $c=1$. At the fixed point we have $\partial_tU(z,\bar{z},t)=\partial_tV(z,\bar{z},t)=0$, which is calculated as follows,
 
\begin{align}
\partial_tV(z,\bar{z},t)=&\frac{\partial^2V(z,\bar{z},t)}{\partial z\partial\bar{z}}-\frac{\partial V(z,\bar{z},t)}{\partial z}\frac{\partial V(z,\bar{z},t)}{\partial\bar{z}}=0\\\nonumber
=&\frac{1}{\lambda t}-\frac{z\bar{z}}{\lambda^2t^2}\\\nonumber
\Rightarrow &t=\frac{z\bar{z}}{\lambda}.
\end{align} 

\noindent Therefore at the fixed point,
\begin{align}
\label{fermion_results_intro}
U\big(z,\bar{z},t=\frac{z\bar{z}}{\lambda}\big)=&e^{1+\ln(z\bar{z})}\\\nonumber
V\big(z,\bar{z},t=\frac{z\bar{z}}{\lambda}\big)=&\big(1+\ln(z\bar{z})\big).
\end{align}

\noindent Notice that the fixed point interaction (vertex) function $V(z,\bar{z},t)$ obtained above for the Fermionic fields in equation \eqref{fermion_results_intro} is also a logarithmic interaction resembling the logarithmic function obtained in equation \eqref{homo_solution} for the Bosonic fields. Which also leads to similar conclusions as in the Bosonic fields case regarding the fixed point being controlled by a logarithmic interaction CFT. Perturbing around the homogeneous solution as follows,

\begin{align}
V_p(z,\bar{z},t)=&V(z,\bar{z},t)+g_f \delta V(z,\bar{z},t)\\\nonumber
=&V(z,\bar{z},t)+g_f e^{\lambda t}f(z,\bar{z}),
\end{align}

\noindent where $g_f\ll 1$ is the Fermion interaction term. At order $\mathcal{O}(g_f)$ ignoring $\mathcal{O}(g_f^2)$ terms we get the following differential equation,

\begin{align}
\label{fer_expansion}
&\frac{\partial^2f(z,\bar{z})}{\partial z\partial\bar{z}}e^{\lambda t}=e^{\lambda t}\bigg[(d-\lambda)f(z,\bar{z})+\frac{\partial f(z,\bar{z})}{\partial z}\frac{\partial V}{\partial\bar{z}}\\\nonumber
&+\frac{\partial V}{\partial z}\frac{\partial f(z,\bar{z})}{\partial\bar{z}}
-\frac{1}{2}(d-1)\bigg(z\frac{\partial f(z,\bar{z})}{\partial z}+\bar{z}\frac{\partial f(z,\bar{z})}{\partial\bar{z}}\bigg)\bigg],
\end{align}

\noindent where,

\begin{align}
\frac{\partial V(z,\bar{z},t)}{\partial \bar{z}}=-\frac{z}{\lambda t},\quad \frac{\partial V(z,\bar{z},t)}{\partial z}=\frac{\bar{z}}{\lambda t}.
\end{align}
\\
\noindent A similar differential equation as the one in equation \eqref{fer_expansion} has been obtained in \cite{harveyfros2001local}. Expanding the exponential $e^{\lambda t}$ on both the LHS and RHS, then collecting the zeroth order terms in parameter $t$ we get the following differential,
\begin{align}
\frac{\partial^2f(z,\bar{z})}{\partial z\partial\bar{z}}=&(d-\lambda)f(z,\bar{z})+\bigg(1-\frac{1}{2}(d-1)\bigg)z\frac{\partial f(z,\bar{z})}{\partial z}\\\nonumber
&+\bigg(1-\frac{1}{2}(d-1)\bigg)\bar{z}\frac{\partial f(z,\bar{z})}{\partial \bar{z}}.
\end{align}
\\
\noindent Performing the following transformation,
\\
\begin{align}
f(z,\bar{z})&\rightarrow \frac{1}{1-\frac{1}{2}(d-1)}f(z^\prime,\bar{z}^\prime)\\\nonumber
&z^\prime =z \sqrt{\bigg(1-\frac{1}{2}(d-1)\bigg)}\\\nonumber
&\bar{z}^\prime = \bar{z}\sqrt{\bigg(1-\frac{1}{2}(d-1)\bigg)},
\end{align} 
\noindent then the differential equation above becomes,
\\
\begin{align}
\frac{\partial^2f(z^\prime,\bar{z}^\prime)}{\partial z^\prime\partial\bar{z}^\prime}=&\frac{(d-\lambda)}{1-\frac{1}{2}(d-1)}f(z^\prime,\bar{z}^\prime)+z^\prime\frac{\partial f(z^\prime,\bar{z}^\prime)}{\partial z^\prime}\\\nonumber
&+\bar{z}^\prime\frac{\partial f(z^\prime,\bar{z}^\prime)}{\partial \bar{z}^\prime}.
\end{align}
\\
\noindent Since $f(z^\prime,\bar{z}^\prime)$ is a Grassmann function, the expansion of the function will terminate at second order in Grassmann variables $z^\prime$ and $\bar{z}^\prime$ because they will be nilpotent. Therefore, the highest degree polynomial function that can be the solution to the differential equation above will be,
\\
\begin{align}
\label{eigen_function1}
f(z^\prime,\bar{z}^\prime)=z^\prime\bar{z}^\prime+\frac{1}{2}.
\end{align} 
\\
\noindent Plugging this eigenfunction/eigenoperator solution to the differential equation above we find the eigenvalue $\lambda=\lambda^{(0)}$ at zeroth order in expansion parameter $t$ to be,
\begin{align}
\lambda^{(0)} = 2\bigg(1-\frac{1}{2}(d-1)\bigg)+d,
\end{align}
\\
\noindent and therefore at $d=4$, $\lambda^{(0)}=3$. At first order in expansion parameter $t$ we get the following differential equation from equation \eqref{fer_expansion},
\\
\begin{align}
\label{diff}
\frac{\partial^2f(z^\prime,\bar{z}^\prime)}{\partial z^\prime\partial\bar{z}^\prime}=&\frac{(d-\lambda)}{\frac{1}{2}-\frac{1}{2}(d-1)}f(z^\prime,\bar{z}^\prime)+z^\prime\frac{\partial f(z^\prime,\bar{z}^\prime)}{\partial z^\prime}\\\nonumber
&+\bar{z}^\prime\frac{\partial f(z^\prime,\bar{z}^\prime)}{\partial \bar{z}^\prime},
\end{align}
\\
\noindent and here we have performed the following transformation,
\\
\begin{align}
f(z,\bar{z})&\rightarrow \frac{1}{\frac{1}{2}-\frac{1}{2}(d-1)}f(z^\prime,\bar{z}^\prime)\\\nonumber
&z^\prime = z\sqrt{\bigg(\frac{1}{2}-\frac{1}{2}(d-1)\bigg)}\\\nonumber
&\bar{z}^\prime =\bar{z}\sqrt{\bigg(\frac{1}{2}-\frac{1}{2}(d-1)\bigg)}.
\end{align} 
\\
\noindent The same eigenfunction given in equation \eqref{eigen_function1} is the solution to the differential equation \eqref{diff}, here we find the corresponding eigenvalue $\lambda=\lambda^{(1)}$ to be,
\\
\begin{align}
\lambda^{(1)} = 2\bigg(\frac{1}{2}-\frac{1}{2}(d-1)\bigg)+d,
\end{align}
\\
\noindent and at $d=4$, $\lambda^{(1)}=2$. Then at the $n^{th}$ order expansion in parameter $t$ we get the following differential equation,
\\
\begin{align}
\frac{\partial^2f(z^\prime,\bar{z}^\prime)}{\partial z^\prime\partial\bar{z}^\prime}=&\frac{(d-\lambda)}{\frac{1}{n}-\frac{1}{2}(d-1)}f(z^\prime,\bar{z}^\prime)+z^\prime\frac{\partial f(z^\prime,\bar{z}^\prime)}{\partial z^\prime}\\\nonumber
&+\bar{z}^\prime\frac{\partial f(z^\prime,\bar{z}^\prime)}{\partial \bar{z}^\prime},
\end{align}
\\
after having performed the following transformation,
\\
\begin{align}
f(z,\bar{z})&\rightarrow \frac{1}{\frac{1}{n}-\frac{1}{2}(d-1)}f(z^\prime,\bar{z}^\prime)\\\nonumber
&z^\prime = z\sqrt{\bigg(\frac{1}{n}-\frac{1}{2}(d-1)\bigg)}\\\nonumber
&\bar{z}^\prime = \bar{z}\sqrt{\bigg(\frac{1}{n}-\frac{1}{2}(d-1)\bigg)}.
\end{align}
\\
The same function already given in equation \eqref{eigen_function1} is the solution here, and the corresponding eigenvalue $\lambda^{(n)}=\lambda$ is 
\begin{align}
\lambda^{(n)} = 2\bigg(\frac{1}{n}-\frac{1}{2}(d-1)\bigg)+d,
\end{align}
\\
\noindent and at $d=4$ when $n\rightarrow\infty$, $\lambda^{(n)}=1$. Take note that since the Grassmann expansion is nilpotent already at the second order in expansion parameter $t$, the expansion terminates at that order, therefore it is unnecessary to go beyond the expansion $n>1$ and $n\rightarrow\infty$ since the terms are zero. Therefore the Fermionic fields expansion in parameter $t$ is finite, unlike in the Bosonic fields where the expansion is infinite. Now the perturbed function $V_p$ for the Fermionic fields will be,

\begin{align}
V_P(z,\bar{z},t)=&\bigg(\frac{z\bar{z}}{\lambda t}+\ln(\lambda t)\bigg)+g_f\sum_n e^{\lambda^{(n)} t}f^{(n)}(z,\bar{z})\\\nonumber
=&\bigg(\frac{z\bar{z}}{\lambda t}+\ln(\lambda t)\bigg)+g_f\bigg(\frac{1}{\lambda t}f^{(-1)}(z,\bar{z})\\\nonumber
&+\lambda^{(0)}f^{(0)}(z,\bar{z})+\lambda^{(1)}tf^{(1)}(z,\bar{z})\bigg)
\end{align}
\noindent where again $f^{(-1)}(z,\bar{z})$ is a constant function, $f^{(-1)}(z,\bar{z})=const$, and $f^{(0)}(z,\bar{z})=f^{(1)}(z,\bar{z})=z\bar{z}-\frac{1}{2}$. Then at the logarithmic fixed point at $t=\frac{z\bar{z}}{\lambda}$, the function $V_p$ becomes,
\begin{align}
V_P(z,\bar{z})=&\big(1+\ln(z\bar{z})\big)+g_f\bigg(\frac{1}{z\bar{z}}f^{(-1)}(z,\bar{z})\\\nonumber
&+\lambda^{(0)}\bigg(z\bar{z}-\frac{1}{2}\bigg)-\frac{1}{2}\frac{\lambda^{(1)}}{\lambda}z\bar{z}\bigg).
\end{align}
\noindent Notice how different the linearized solution for the Fermionic fields  is different from the Bosonic fields solution in equation \eqref{b_results}. First, the expansion terminates after the first order in expansion parameter $t$ while for the Bosonic fields the expansion is infinite, and second, the polynomial function is not a Hermite polynomial which has $n^{th}$ degree expansion as it was in the case of the Bosonic fields, it is a simple quadratic function $f(z,\bar{z})$ with no higher degree expansion terms. The truncation in the expansion is due to the nipotent property of Grassmann variables representing the Fermionic fields.\\
\\
Towards this end, in $d=4$, at zeroth order in expansion parameter $t$ we find the correlation length critical exponent is $\nu=0.33$, and at first order in expansion parameter  $t$ the correlation length critical exponent $\nu=0.5$. From here the expansion terminates since we are dealing with Grassmann variables, and therefore there will be no other higher order critical exponents in expansion parameter $t$. Furthermore, there are no negative eigenvalues which indicates the absence of correction-to-scaling critical exponents.

\section{Discussion}

\noindent In the paper we have demonstrated that the Polchinski's ERG flow equation given in \cite{POLCHINSKI1984269}, within the LPA limit it is equivalent to the standard elementary heat equation. The fixed points (steady-state) of this equation are controlled by a logarithmic function which is the interaction function of the theory, we therefore conclude that this limit gives rise to a logarithmic interaction CFT. Contrary to perturbing (linearizing) around the Gaussian fixed point, here we perturb around the logarithmic fixed point since it is the solution of our steady-state equation. For the case of the Bosonic fields the perturbed solution leads to an finite set of Hermite differentials with Hermite polynomials being the solutions to these differentials. For the Fermionic fields case we find a finite set of Grassmann differentials analogue to Hermite differentials, the set is finite because the expansion truncates at the second order in expansion parameter $t$, since the Grassmann variables are nilpotent at this order. Then we proceeded to compute the relevant critical exponents for both Bosonic fields and Fermionic fields cases; For the Bosonic fields we end up with an infinite set of critical exponents since the values of the critical exponent are different at each order in expansion parameter $t$ due to the relevant eigenvalues being different at each order; For the Fermionic fields we end up with a finite set of critical exponents and the critical exponents are also different for the zero and first order expansion in parameter $t$, and the other higher order expansions are nilpotent. Furthermore, there are no negative eigenvalues for Fermionic fields case, and therefore the correction-to-scaling critical exponents are absent.\\
\\
\noindent The simplistic results obtained at the LPA limit we find them valuable because it opens up a new avenue to possible study and understand ERG equations through using well established tools and theories already being utilized to solve and understand heat equations. Furthermore these results deserve an in-depth probing, at this point there are still some open questions on whether this relation at the LPA limit is something of a mere coincidence or is a smoking gun to something fundamental. One way to address such questions will be to move the computation beyond the LPA limit by including higher order terms in the derivative expansion. However, presently the challenge with including higher order terms in derivative expansion or momentum expansion has been seen to result with physical observables being dependent on the choices of the renormalization scheme \cite{Ball_1995,Morris_1994,Tetradis_1994}. Meaning the theory will be scheme dependent, and the RG flow equations will depend on the conditions of the theory, these conditions will include the choices used in the renormalization scheme to fix the theory. This makes the renormalization of next to leading order and higher order derivative terms tricky to deal with, and can yield insensible physical observable results. Since in any renormalization scheme there are always some choices that one has to make, and these choices are related to the renormalization conditions that one implements on the theory, and these choices have to be fixed. In a sensible renormalization scheme, the physical observables computed from this theory are suppose to be renormalization scheme independent, independent of the renormalization conditions. For instance, in perturbation renormalization group for on-shell renormalization scheme, one of the choices is the propagator has a pole at mass squared with residue one. Another drawback with this ERG approximation approach whereby we perform derivative or momentum expansion where at some point we truncate the expansion is that the expansion is not systematic, one can't estimate the error they make when they truncate the expansion, therefore better systematic truncation approaches still require investigation here. Including higher order derivative terms is one of the immediate interesting direction to extend this work. Going forward, we also need to make contact with experimental results, search for relevant experimental results that can be used to verify our computational results.

\section*{Acknowledgements}
\noindent The author would like to thank Robert De Mello Koch for suggesting the project and providing valuable discussions and comments.

\begin{widetext}
\section*{A\quad Polchinski's ERG flow equation for Bosonic and Fermionic fields}

\noindent We used section 2 to formulate Polchincki's ERG flow equation for both Fermionic and Bosonic fields, and we have written down two types of Polchinski's ERG flow equation, the first one in equation \eqref{general_polchinski} only implements the coarse graining step of Wilson's renormalization group flow program, and the second one in equation \eqref{integro} implements both the coarse-graining and rescaling steps of Wilson's renormalization group flow program. This section will provide more details regarding the derivations of these equations.
\subsection*{A.1 Polchinski's ERG flow equation coarse-graining step}
\noindent At the start of section 2 and particularly in equation \eqref{general_polchinski} we have shown how the Polchinski's ERG flow equation containing both the single Bosonic fields and the spinless Fermionic fields can be formulated. The detailed derivation for the formulation is as follows,

\begin{align}
&e^{-W_{\Lambda}^{\Lambda_0}[J]}=\int D\psi D\bar{\psi}D\phi e^{-\frac{1}{2}\int_p\phi(p)G_B^{-1}(\frac{p}{\Lambda})\phi(-p)-\int_p\bar{\psi}(p)G_F^{-1}(\frac{p}{\Lambda})\psi(-p)+\int_pJ(p)\phi(-p)+\int_p\bar{\eta}(p)\psi(-p)+\int_p\eta(p)\bar{\psi}(-p)-S_{int}[\phi,\bar{\psi},\psi]}.\\\nonumber
\end{align}
\noindent Applying the equation below on the Bosonic fields,
\begin{align}
\label{aabb}
e^{-W_{\Lambda}^{\Lambda_0}[J]}=&\int D\phi e^{-\int_p\frac{1}{2}\phi(p)G^{-1}_B(\frac{p}{\Lambda})\phi(-p)+\int_pJ(p)\phi(-p)-S_{int}^{\Lambda_0}[\phi]}\\\nonumber
=&e^{-\frac{1}{2}\int_pJ(p)G_B(\frac{p}{\Lambda})J(-p)}e^{D_{\Lambda}^{\Lambda_0}}e^{-S_{int}^{\Lambda_0}[\phi]}\bigg|_{\phi=G_{B}J(p)},
\end{align}

\noindent where

\begin{align}
D_{\Lambda}^{\Lambda_0}=&\frac{1}{2}\int_pG_B\bigg(\frac{p}{\Lambda}\bigg)\frac{\delta}{\delta\phi(p)}\frac{\delta}{\delta\phi(-p)}\\\nonumber
=&\frac{1}{2}\int\frac{d^dp}{(2\pi)^d}G_B\bigg(\frac{p}{\Lambda}\bigg)\frac{\delta}{\delta\phi(p)}\frac{\delta}{\delta\phi(-p)},
\end{align}

\noindent we get
\begin{align}
e^{-W_{\Lambda}^{\Lambda_0}[J]}=&\int D\psi D\bar{\psi}e^{-\int_p\bar{\psi}(p)G_F^{-1}(\frac{p}{\Lambda})\psi(-p)+\int_p\bar{\eta}(p)\psi(-p)+\int_p\eta(p)\bar{\psi}(-p)}\\\nonumber
&\times e^{-\frac{1}{2}\int_pJ(p)G_B(\frac{p}{\Lambda})J(-p)}e^{\frac{1}{2}\int_pG_B(\frac{p}{\Lambda})\frac{\delta}{\delta\phi(p)}\frac{\delta}{\delta\phi(-p)}e^{-S_{int}[\phi,\bar{\psi},\psi]}}\bigg|_{\phi=G_B(\frac{p}{\Lambda})J(p)}\\\nonumber
=&e^{-\frac{1}{2}\int_pJ(p)G_B(\frac{p}{\Lambda})J(-p)}\int D\psi D\bar{\psi}e^{-\int_p\bar{\psi}(p)G_F^{-1}(\frac{p}{\Lambda})\psi(-p)+\int_p\bar{\eta}(p)\psi(-p)+\int_p\eta(p)\bar{\psi}(-p)}\\\nonumber
&\times e^{\frac{1}{2}\int_pG_B(\frac{p}{\Lambda})\frac{\delta}{\delta\phi(p)}\frac{\delta}{\delta\phi(-p)}e^{-S_{int}[\phi,\bar{\psi},\psi]}}\bigg|_{\phi=G_B(\frac{p}{\Lambda})J(p)}.\\\nonumber
\end{align}

\noindent Then applying equation \eqref{aabb} on the Fermionic fields we get

\begin{align}
e^{-W_{\Lambda}^{\Lambda_0}[J]}=&e^{-\frac{1}{2}\int_pJ(p)G_B(\frac{p}{\Lambda})J(-p)}\bigg[e^{-\int_p\bar{\eta}(p)G_F\eta(-p)}e^{-\int_pG_F\frac{\delta}{\delta\bar{\psi}(p)}\frac{\delta}{\delta\psi(-p)}}e^{-\frac{1}{2}\int_pG_B(\frac{p}{\Lambda})\frac{\delta}{\delta\phi(p)}\frac{\delta}{\delta\phi(-p)}}e^{-S_{int}[\phi,\bar{\psi},\psi]}\bigg]\bigg|_{\phi=G_BJ,\bar{\psi}=G_F\eta,\psi=G_F\bar{\eta}}
\end{align}

\noindent Defining the effective interaction action $S_{int}[\phi,\bar{\psi},\psi,\Lambda]$ from the functional integral above as,

\begin{align}
e^{-S_{int}[\phi,\bar{\psi},\psi,\Lambda]} = e^{-\frac{1}{2}\int_pJ(p)G_B(\frac{p}{\Lambda})J(-p)}e^{-\int_p\bar{\eta}(p)G_F(\frac{p}{\Lambda})\eta(-p)}e^{-\int_pG_F(\frac{p}{\Lambda})\frac{\delta}{\delta\bar{\psi}(p)}\frac{\delta}{\delta\psi(-p)}}e^{-\frac{1}{2}\int_pG_B(\frac{p}{\Lambda})\frac{\delta}{\delta\phi(p)}\frac{\delta}{\delta\phi(-p)}}e^{-S_{int}[\phi,\bar{\psi},\psi]},
\end{align}

\noindent taking the derivative with respect the scale $\Lambda$ we end up with the following Polchinski's ERG flow equation for both Bosonic fields and Fermionic fields,

\begin{align}
\label{appendix_renorm}
&-\Lambda\frac{\partial S_{int}}{\partial\Lambda}=\int_{p}\Lambda\frac{\partial G_F(\frac{p}{\Lambda})}{\partial\Lambda}\bigg(\frac{\delta S_{int}}{\delta\bar{\psi}(p)}\frac{\delta S_{int}}{\delta\psi(-p)}-\frac{\delta^2S_{int}}{\delta\bar{\psi}(p)\delta\psi(-p)}\bigg)+\frac{1}{2}\int_{p}\Lambda\frac{\partial G_B(\frac{p}{\Lambda})}{\partial\Lambda}\bigg(\frac{\delta S_{int}}{\delta\phi(p)}\frac{\delta S_{int}}{\delta\phi(-p)}-\frac{\delta^2S_{int}}{\delta\phi(p)\delta\phi(-p)}\bigg),
\end{align}

\noindent which is one of the main equations we were trying to derive in section 2. This part of the derivation only considers the coarse graining step of Wilson's renormalization group program. In the next part we show how the second step of Wilson's renormalization group flow is implemented on this equation \eqref{appendix_renorm}, this step entails rescaling the Polchinski's ERG flow equation's parameters (variables).

\subsection*{A.2 Rescaling Polchinski's ERG flow equation.}
\noindent At the beginning of section 2 the Polchinski's ERG flow equation derived earlier only implements the coarse step, therefore the rescaling step is required in order to complete Wilson's renormalization group flow program. We will show how this step is implemented which will lead to the derivation of equation \eqref{integro} shown at the end of section 2. First, the full rescaled Polchinski's ERG flow equation derived in \cite{Bagnuls_2001,Rosten_2012} for the single Bosonic fields is as follows,

\begin{align}
\bigg(\Lambda\frac{\partial}{\partial\Lambda}+\int\frac{d^dp}{(2\pi)^d}\phi(p)p\cdot\frac{\partial}{\partial p}\frac{\delta}{\delta\phi(-p)}+d_{\phi}\int_p\phi_p\frac{\delta}{\delta\phi_p}-d\bigg)S[\phi,\Lambda]=\frac{1}{2}\int_{p}\Lambda\frac{\partial G_B(\frac{p}{\Lambda})}{\partial\Lambda}\bigg(\frac{\delta S[\phi,\Lambda]}{\delta\phi(p)}\frac{\delta S[\phi,\Lambda]}{\delta\phi(-p)}-\frac{\delta^2S[\phi,\Lambda]}{\delta\phi(p)\delta\phi(-p)}\bigg)
\end{align}

\noindent where $d_{\phi}=\frac{1}{2}(d-2+\eta)$, $\eta$ is the anomalous dimension of the scalar fields, $d$ is the space-time dimension, and take note that $\int_p=\int\frac{d^dp}{(2\pi)^d}$. Here we will follow similar arguments used in \cite{Bagnuls_2001, Rosten_2012} to apply rescaling on Polchinski's ERG flow equation. Dropping the Fermionic fields from the effective action $S_{int}$ in equation \eqref{general_int}, we get the following effective action,

\begin{align}
\label{no_fermions}
S[\phi,\Lambda]=\sum_{m}\int\prod_{i=1}^m\frac{d^dp_i}{(2\pi)^d}\mathcal{L}_{m,0}(p_1,\cdots ,p_m,\Lambda)\phi(p_1)\cdots\phi(p_m)\delta^d(p_1+\cdots +p_m).
\end{align}

\noindent where $S[\phi,\Lambda]=S_{int}[\phi,0,0,\Lambda]$. Now, rescaling the momenta $p$ as,

\begin{align}
p\rightarrow p^\prime = sp = (1+t)p
\end{align}
\noindent where $s=e^{t}$ and $t=\frac{\delta\Lambda}{\Lambda}$ is an infinitesimal small parameter. The rescaling has consequences on the action $S=S[\phi,\Lambda]$, it transforms the action $S$ in the following way,
\begin{align}
S\rightarrow S^\prime =& S+\Delta S\\\nonumber
=&S+tG_{sc}S\\\nonumber
\Rightarrow \frac{\partial S}{\partial t}=&G_{sc}S
\end{align}

\noindent where the factor $\Delta S= t G_{sc}S$ and $G_{sc}$ will contain correction terms to the action $S$ that arise from rescaling. The factor $\Delta S = t G_{sc}S$ is actually a sum from various factors or terms arising from inducing changes in the action of equation \eqref{no_fermions} and in the end these terms are summed together as $\Delta S=\sum_{i=1}^4\Delta S_i$. We can unpack $\Delta S_1$ as coming from the differential volume,
\begin{align}
\int\prod_{i=1}^m d^dp_1\cdots d^dp_m = \int\prod_{i=1}^m(1+t)^{-md}d^dp^\prime_1\cdots d^dp^\prime_m
\end{align}
\noindent expanding to first order in parameter $t$ this induces $\Delta S_1$ as,
\begin{align}
\Delta S_1 = -t\bigg(d\int_p\phi_p\frac{\delta}{\delta\phi_p}\bigg)S.
\end{align}

\noindent The factor $\Delta S_2$ is induced from the field rescaling $\phi(p)=\phi(s^{-1}p^\prime)$, giving,

\begin{align}
\phi(s^{-1}p^\prime)=(1+t)^{d-d_\phi}\phi(p),
\end{align}
\noindent to first order in parameter $t$ the inducing factor is,
\begin{align}
\Delta S_2 = t\bigg((d-d_{\phi})\int_p\phi_p\frac{\delta}{\delta\phi_p}\bigg)S,
\end{align}

\noindent where $d_{\phi}=\frac{1}{2}(d-2+\eta_{\phi})$ and $\eta_{\phi}$ is the anomalous dimension of the scalar field $\phi(p)$. The factor $\Delta S_3$ is induced from the interaction or vertex function $\mathcal{L}_{m,0}(p_1,\cdots,p_m,\Lambda)=\mathcal{L}_{m,0}(s^{-1}p_1^\prime,\cdots,s^{-1}p_m^\prime;\Lambda)$, which gives,

\begin{align}
\Delta S_3 = -t\bigg(\int_p\phi_p p\cdot\frac{\partial}{ \partial p^\prime} \frac{\delta}{\delta\phi_p}\bigg)S.
\end{align}

\noindent Lastly, the term $\Delta S_4$ is induced from the delta function $\delta^d(p_1+\cdots+p_m)=\delta^d(s^{-1}p_1^\prime+\cdots+s^{-1}p_m^\prime)$, giving
\begin{align}
\Delta S_4 = t(dS).
\end{align}
\noindent Then in the end these induced terms from rescaling are now summed together,
\begin{align}
\Delta S =& \Delta S_1+\Delta S_2+\Delta S_3+\Delta S_4\\\nonumber
=&t\bigg(d-\int_p\phi_p p\cdot \partial_{p^\prime}\frac{\delta}{\delta\phi_p}-d_{\phi}\int_p\phi_p\frac{\delta}{\delta\phi_p}\bigg)S[\phi,\Lambda].
\end{align}
\noindent Allowing the momentum derivative $\partial_{p^\prime}$ to act on the momentum conservation function of $S[\phi,\Lambda]$ we get,
\begin{align}
\Delta S = -t\bigg(\int_p\phi_p p\cdot \partial_{p}\frac{\delta}{\delta\phi_p}+d_{\phi}\int_p\phi_p\frac{\delta}{\delta\phi_p}\bigg)S[\phi,\Lambda],
\end{align}

\noindent then,
\begin{align}
G_{sc}S[\phi,\Lambda]=&-\bigg(\int_p\phi_p p\cdot \partial_{p}\frac{\delta}{\delta\phi_p}+d_{\phi}\int_p\phi_p\frac{\delta}{\delta\phi_p}\bigg)S[\phi,\Lambda]\\\nonumber
=&(d-\Delta_{\partial_{\phi}}-d_{\phi}\Delta_{\phi})S[\phi,\Lambda]\\\nonumber
\Rightarrow G_{sc}=&(d-\Delta_{\partial_{\phi}}-d_{\phi}\Delta_{\phi}),
\end{align}

\noindent where $\Delta_{\partial_{\phi}}=d+\int_p\phi_pp\cdot \partial_{p}\frac{\delta}{\delta\phi_p}$, $\Delta_{\phi}=\int_p\phi_p\frac{\delta}{\delta\phi_p}$ and $d_{\phi}=\frac{1}{2}(d-2+\eta_{\phi})$. Therefore, after applying rescaling on the single Bosonic fields we get the following Polchinski's ERG flow equation,

\begin{align}
\bigg(\Lambda\partial_{\Lambda}+d_\phi \bigtriangleup_{\phi}+\bigtriangleup_{\partial_\phi}-d\bigg)S[\phi,\Lambda] =
&\frac{1}{2}\int_{p} G_B^\prime\big(\frac{p}{\Lambda}\big)\bigg(\frac{\delta S[\phi,\Lambda]}{\delta\phi(p)}\frac{\delta S[\phi,\Lambda]}{\delta\phi(-p)}-\frac{\delta^2S[\phi,\Lambda]}{\delta\phi(p)\delta\phi(-p)}\bigg)\\\nonumber
\bigg(\partial_{t}+d_\phi \bigtriangleup_{\phi}+\bigtriangleup_{\partial_\phi}-d\bigg)S[\phi,t] =
&\frac{1}{2}\int_{p} G_B^\prime\big(\frac{p}{\Lambda}\big)\bigg(\frac{\delta S[\phi,t]}{\delta\phi(p)}\frac{\delta S[\phi,t]}{\delta\phi(-p)}-\frac{\delta^2S[\phi,t]}{\delta\phi(p)\delta\phi(-p)}\bigg),
\end{align}

\noindent where in the last line we used the fact that $\Lambda\frac{\partial}{\partial\Lambda}=\frac{\partial}{\partial t}$. Now, it should be easy to infer that the inclusion of Fermionic fields back into the effective action $S_{int}$ in equation \eqref{general_int} while removing the Bosonic fields at the same time, then applying rescaling should have the following effect on the spinless Fermionic fields effective action,
\begin{align}
\Delta S =& t\bigg(\int_pd_{\psi\bar{\psi}}\psi(p)\frac{\delta}{\delta\psi(-p)}+\int_pd_{\psi\bar{\psi}}\bar{\psi}(p)\frac{\delta}{\delta\bar{\psi}(-p)}+\int_p\psi(p)p\cdot\frac{\partial}{\partial p}\frac{\delta}{\delta\psi(-p)}+\int_p\bar{\psi}(p)p^\cdot\frac{\partial}{\partial p}\frac{\delta}{\delta\bar{\psi}(-p)}\bigg)S[\psi,\bar{\psi},\Lambda]
\end{align}

\noindent where $d_{\psi\bar{\psi}}=\frac{1}{2}(d-1+\eta_{\psi})$ and $\eta_{\psi}$ is the anomalous dimensions of the Fermionic fields. We can continue to write $\Delta S$ as,

\begin{align}
\Delta S=&-t\bigg(-d+d_{\psi\bar{\psi}}\Delta_{\psi\bar{\psi}}+\Delta_{\partial_{\psi\bar{\psi}}}\bigg)S[\psi,\bar{\psi},\Lambda]
\end{align}
\noindent where $\Delta_{\psi\bar{\psi}}=\int_p\psi(p)\frac{\delta}{\delta\psi(-p)}+\int_p\bar{\psi}(p)\frac{\delta}{\delta\bar{\psi}(-p)}$  and $\Delta_{\partial_{\psi\bar{\psi}}}=d+\int\frac{d^dp}{(2\pi)^d}\psi(p)p\cdot\frac{\partial}{\partial p}\frac{\delta}{\delta\psi(-p)}+\int\frac{d^dp}{(2\pi)^d}\bar{\psi}(p)p^\cdot\frac{\partial}{\partial p}\frac{\delta}{\delta\bar{\psi}(-p)}$. Then the factor $G_{sc}$ from rescaling the Fermionic action will be

\begin{align}
G_{sc}S[\psi,\bar{\psi},\Lambda] = \bigg(d-d_{\psi\bar{\psi}}\Delta_{\psi\bar{\psi}}-\Delta_{\partial_{\psi\bar{\psi}}}\bigg)S[\psi,\bar{\psi},\Lambda].
\end{align}
\noindent

\noindent Therefore, the rescaled Polchinski's ERG flow equation for the spinless Fermionic fields is,

\begin{align}
\bigg(\partial_t-d+d_{\psi\bar{\psi}}\Delta_{\psi\bar{\psi}}+\Delta_{\partial_{\psi\bar{\psi}}}\bigg)S[\psi,\bar{\psi},t]=\int_{p} G_F^\prime\big(\frac{p}{\Lambda}\big)\bigg(\frac{\delta S[\psi,\bar{\psi},t]}{\delta\psi(p)}\frac{\delta S[\psi,\bar{\psi},t]}{\delta\bar{\psi}(-p)}-\frac{\delta^2S[\psi,\bar{\psi},t]}{\delta\psi(p)\delta\bar{\psi}(-p)}\bigg).
\end{align}
\\
\noindent Now, combining the rescaled Polchinski's ERG flow equations for the single Bosonic and spinless Fermionic fields together we get the following rescaled ERG flow equation,
\\
\begin{align}
\big(\partial_t+d_{\psi\bar{\psi}}\Delta_{\psi\bar{\psi}}+\Delta_{\partial_{\psi\bar{\psi}}}+&d_\phi \bigtriangleup_{\phi}+\bigtriangleup_{\partial_\phi}-d\big)S[\phi, \psi,\bar{\psi},t]=\int_{p} G_F^\prime\big(\frac{p}{\Lambda}\big)\bigg(\frac{\delta S[\phi,\psi,\bar{\psi},t]}{\delta\psi(p)}\frac{\delta S[\phi,\psi,\bar{\psi},t]}{\delta\bar{\psi}(-p)}-\frac{\delta^2S[\phi,\psi,\bar{\psi},t]}{\delta\psi(p)\delta\bar{\psi}(-p)}\bigg)\\\nonumber
&+\frac{1}{2}\int_{p} G_B^\prime\big(\frac{p}{\Lambda}\big)\bigg(\frac{\delta S[\phi,\psi,\bar{\psi},t]}{\delta\phi(p)}\frac{\delta S[\phi,\psi,\bar{\psi},t]}{\delta\phi(-p)}-\frac{\delta^2S[\phi,\psi,\bar{\psi},t]}{\delta\phi(p)\delta\phi(-p)}\bigg).
\end{align}

\section*{B\quad Local Potential Approximation}
\noindent In section 3 we approximate non-perturbatively the Polchinski's ERG flow equation by performing derivative expansions, which is an expansion in powers of momenta. The expansion is truncated at the LPA by making use of the Hazenfratz and Hazenfratz projector (operator), which extracts the LPA piece of the expansion.

\subsection*{B.1 Bosonic fields Local Potential Approximation (LPA).}
\noindent The LPA derived in section 3 is the leading term or zeroth order term in the derivative expansion of the action $S[\phi,\Lambda]$. The action $S[\phi,\Lambda]$ in momentum space has the following expansion in momentum $k^2$,

\begin{align}
S[\phi,\Lambda] =& \sum_{m}\int\prod_{i=1}^m\frac{d^dk_i}{(2\pi)^d}\bigg( \mathcal{L}_{m,0}+\frac{k_1^2+\cdots+k_m^2}{m(m-1)}\mathcal{L}_{m-2,0}+\mathcal{O}(k^4)+\cdots \bigg)\phi(k_1)\cdots\phi(k_m)\delta^d(k_1+\cdots +k_m)\\\nonumber
=&\int\prod_{i=1}^m\frac{d^dk_i}{(2\pi)^d}\bigg(V(\phi,\Lambda)+W(\phi,\Lambda)+\mathcal{O}(k^4)\bigg)
\end{align}
\noindent where $V(\phi,\Lambda)=\mathcal{L}_{m,0}(0,\cdots 0;\Lambda)\phi(k_1)\cdots\phi(k_m)\delta^d(k_1+\cdots +k_m)$ is a local vertex or interaction term, while the $W(\phi,\Lambda)$ and other higher order expansion terms are non-local vertex or coupling functions. In position space this action is written as,

\begin{align}
S[\phi,\Lambda]=\int d^dx\bigg(V(\phi,\Lambda)+W(\phi,\Lambda)\partial_{\mu}\phi\partial^{\mu}\phi +\mathcal{O}(\partial^4)\bigg).
\end{align}

\noindent We can focus on the local potential approximation limit by removing or neglecting terms from $S[\phi,\Lambda]$ which are non-local or contain derivatives on them. This can be achieved by making use of the Hasenfratz and Hasenfratz operator $P(z)$        \cite{Hasenfratz:1985dm}. Where for example the action of this operator on a general test functional field $X[\phi]$ acts as follows,

\begin{align}
\label{boson_hazen}
P(z)X[\phi]=e^{z\frac{\partial}{\partial\phi(0)}}X[\phi]\bigg|_{\phi=0}.
\end{align}

\noindent Then on the product of test functionals $X_1[\phi]X_2[\phi]\cdots X_m[\phi]$, the action of the operator $P(z)$ simplifies as follows,

\begin{align}
P(z)\bigg(X_1[\phi]X_2[\phi]\cdots X_m[\phi]\bigg)=&e^{z\frac{\partial}{\partial\phi(0)}}\bigg(X_1[\phi]X_2[\phi]\cdots X_m[\phi]\bigg)\\\nonumber
=&P(z)X_1[\phi]P(z)X_2[\phi]\cdots P(z)X_m[\phi].
\end{align}

\noindent Now using this simplification, the action of the operator $P(z)$ on $S[\phi,\Lambda]$ yields the following results,

\begin{align}
\label{r1}
P(z)S[\phi,\Lambda]=&e^{z\frac{\partial}{\partial\phi(0)}}S[\phi,\Lambda]\bigg|_{\phi=0}\\\nonumber
=&e^{z\frac{\partial}{\partial\phi(0)}}\bigg(\sum_{m}\int\prod_{i=1}^m\frac{d^dk_i}{(2\pi)^d}\mathcal{L}_{m,0}(k_1,\cdots,k_m,\Lambda)\phi(k_1)\cdots\phi(k_m)\delta^d(\sum_i^mk_i)\bigg)\bigg|_{\phi=0}\\\nonumber
=&\sum_m\mathcal{L}_{m,0}(0,0,\cdots ,0,\Lambda)z^m\delta^d(0)\\\nonumber
=&S[z,\Lambda]\delta^d(0)\\\nonumber
=&V(z,\Lambda)\delta^d(0)
\end{align}

\noindent where $S[z,\Lambda]=V(z,\Lambda)=\sum_m\mathcal{L}_{m,0}(0,0,\cdots,0,\Lambda)z^m$. Therefore, after applying the operator $P(z)$ we see that the fields $\phi$ are replaced by the variable $z$, and the action $S[z,\Lambda]$ contains only the local terms or local potential terms which are denoted by the function $V(z,\Lambda)$. Following \cite{Hasenfratz:1985dm}, the delta function $\delta^d(0)$ is handled by working in a finite volume box. Going back to the full rescaled Polchinski's ERG flow equation containing only Bosonic fields, which is

\begin{align}
\bigg(\partial_t+d_{\phi}\triangle_{\phi}+\triangle_{\partial}-d\bigg)S[\phi,t]= \int_{p} G_B^\prime\big(\frac{p}{\Lambda}\big)\bigg(\frac{\delta S[\phi,t]}{\delta\bar{\phi}(p)}\frac{\delta S[\phi,t]}{\delta\phi(-p)}-\frac{\delta^2S[\phi,t]}{\delta\bar{\phi}(p)\delta\phi(-p)}\bigg).
\end{align}

\noindent First take note that the action of the operator $P(z)$ on the term $\frac{\delta S[\phi,t]}{\delta \phi}$ yields the following results,

\begin{align}
\label{r2}
\int_pP(z)\frac{\delta S[\phi,t]}{\delta\phi(p)}\bigg|_{\phi=0}=&\int_pe^{z\frac{\partial}{\partial\phi(0)}}\frac{\delta}{\delta\phi(p)}\bigg(\sum_{m}\int\prod_{i=1}^m\frac{d^dk_i}{(2\pi)^d}\mathcal{L}_{m,0}(k_1,\cdots,k_m,t)\phi(k_1)\cdots\phi(k_m)\delta^d(\sum_i^mk_i)\bigg)\bigg|_{\phi=0}\\\nonumber
=&\sum_m\mathcal{L}_{m,0}(p,0,\cdots 0;\Lambda)mz^{m-1}\delta^d(p)\\\nonumber
=&V^\prime(z,t)\delta^d(p),
\end{align}
\noindent where $V(z,t)$ is a local potential and $V^\prime(z,t)$ is the derivative of the function $V(z,t)$ with respect to $z$. From here one can show the following,

\begin{align}
\label{r3}
\int_p G_B^\prime\big(\frac{p}{\Lambda}\big)\bigg(P(z)\frac{\delta S[\phi,t]}{\delta\phi_p}\bigg)&\bigg(P(z)\frac{\delta S[\phi,t]}{\delta\phi_{-p}}\bigg)=\int_p G_B^\prime\big(\frac{p}{\Lambda}\big)\bigg(\sum_m\mathcal{L}_{m,0}(k_1,\cdots,k_m;t)mz^{m-1}\delta^d(p)\bigg)\\\nonumber
&\times\bigg(\sum_n\mathcal{L}_{n,0}(k_1,\cdots,k_m;t)nz^{n-1}\delta^d(-p)\bigg)\\\nonumber
=& G_B^\prime(0)\bigg(\sum_m\mathcal{L}_{m,0}(0,\cdots,0;t)mz^{m-1}\bigg)\bigg(\sum_n\mathcal{L}_{n,0}(0,\cdots,0;t)nz^{n-1}\bigg)\delta^d(0)\\\nonumber
=& G_B^\prime(0)V^\prime(z,t)V^\prime(z,t)\delta^d(0)\\\nonumber
=& G_B^\prime(0)V^{\prime 2}(z,t)\delta^d(0),
\end{align}

\noindent and also,

\begin{align}
\label{r4}
P(z)\bigg(\int_p G_B^\prime\big(\frac{p}{\Lambda}\big)\frac{\delta^2S[\phi,t]}{\delta\phi_p\delta\phi_{-p}}\bigg)=&\int_p G_B^\prime\big(\frac{p}{\Lambda}\big)\sum_m\mathcal{L}_{m,0}(p,-p,0,\cdots,0;t)m(m-1)z^{m-2}\delta^d(p-p)\\\nonumber
=&\int_p G_B^\prime\big(\frac{p}{\Lambda}\big)\sum_m\mathcal{L}_{m,0}(p,-p,0,\cdots,0;t)m(m-1)z^{m-2}\delta^d(0)\\\nonumber
=&\int_p G_B^\prime\big(\frac{p}{\Lambda}\big)V^{\prime\prime}(z,t)\delta^d(0)\\\nonumber
=&V^{\prime\prime}(z,t) G_B^\prime\big(\frac{p}{\Lambda}\big)\delta^d(0).
\end{align}

\noindent Now when the operator $P(z)$ acts on the entire rescaled Polchinski's ERG flow equation as follows,

\begin{align}
P(z)\bigg(\partial_t+d_{\phi}\triangle_{\phi}+\triangle_{\partial}-d\bigg)S[\phi,t]=P(z)\bigg(\int_{p} G_B^\prime\big(\frac{p}{\Lambda}\big)\bigg(\frac{\delta S[\phi,t]}{\delta\bar{\phi}(p)}\frac{\delta S[\phi,t]}{\delta\phi(-p)}-\frac{\delta^2S[\phi,t]}{\delta\bar{\phi}(p)\delta\phi(-p)}\bigg)\bigg),
\end{align}

\noindent plugging the results in equations \eqref{r1}, \eqref{r2}, \eqref{r3} and \eqref{r4} we obtain the following results,

\begin{align}
\bigg(\partial_t+d_{\phi}\triangle_{\phi}+\triangle_{\partial}-d\bigg)V(z,t)= G_B^\prime(0)V^{\prime 2}(z,t)-V^{\prime\prime}(z,t)\int_p G_B^\prime\big(\frac{p}{\Lambda}\big)
\end{align}

\noindent where now $d_{\phi}=\frac{d-2}{2}$ and the anomalous dimension $\eta_\phi =0$ in the LPA limit since there is no momentum dependence in the zeroth order of the expansion. Therefore in the end we end up with the following differential equation,
\begin{align}
\partial_tV(z,t)=&\int_p G_B^\prime\big(\frac{p}{\Lambda}\big)V^{\prime\prime}(z,t)- G_B^\prime(0)V^{\prime 2}(z,t)-d_{\phi}zV^{\prime}(z,t)+dV(z,t)\\\noindent
=&V^{\prime\prime}(z,t)-V^{\prime 2}(z,t)-d_{\phi}zV^{\prime}(z,t)+dV(z,t),
\end{align}

\noindent after writing $V(z,t)$ and $z$ as

\begin{align}
V(z,t)\rightarrow& \frac{V(z,t)\int_pG_B^\prime(\frac{p}{\Lambda})}{G_B^\prime(0)},\\\nonumber
z\rightarrow& z\sqrt{\int_pG_B^\prime\big(\frac{p}{\Lambda}\big)}.\\\nonumber
\end{align}

\subsection*{B.2 Set of Hermite differential equations}
\noindent In section 3 after obtaining the solution in equation \eqref{homo_sol} which subsequently at the fixed point the solution is a logarithmic function, perturbing close to this solution we get an infinite set tower of Hermite differential equations at each order in expansion parameter $t$. Plugging $V_p(z,t)$ into the full rescaled Polchinski's differential equation at the LPA limit in equation \eqref{rescaled_pol},
\begin{align}
\partial_tV_p(z,t)= V^{\prime\prime}_p(z,t)-V^{\prime 2}_p(z,t)-d_{\phi}zV^{\prime}_p(z,t)+dV_p(z,t),
\end{align}

\noindent where $V_p(z,t)=V(z,t)+g_b\sum_nf_n(z)e^{\lambda_n t}$. Expanding and unpacking the differential equation above we get, 

\begin{align}
\label{epsilon-exp}
&\partial_tV(z,t)+g_b\sum_n\lambda_nf_n(z)e^{\lambda_n t}=\bigg(\frac{\partial^2V(z,t)}{\partial z\partial z}+g_b\sum_n\frac{\partial^2f_n(z)}{\partial z\partial z}e^{\lambda_n t}\bigg)\\\nonumber
&-\bigg(\bigg(\frac{\partial V(z,t)}{\partial z}\bigg)^2+2g_b\frac{\partial V(z,t)}{\partial z}\sum_n\frac{\partial f_n(z)}{\partial z}e^{\lambda_n t}+g^2_b\sum_{m,n}\frac{\partial f_n(z)}{\partial z}\frac{\partial f_m(z)}{\partial z}e^{(\lambda_n+\lambda_m)t}\bigg)\\\nonumber
&-d_{\phi}\bigg(z\frac{\partial V(z,t)}{\partial z}+zg_b\sum_n\frac{\partial f_n(z)}{\partial z}e^{\lambda_nt}\bigg)+d\bigg(V(z,t)+g_b\sum_nf_n(z)e^{\lambda_nt}\bigg)\\\nonumber
&\underbrace{\partial_tV(z,t)-\frac{\partial^2V(z,t)}{\partial z\partial z}+\bigg(\frac{\partial V(z,t)}{\partial z}\bigg)^2}_{=0,\text{ from homogeneous equation}}=-g_b\sum_n\lambda_nf_n(z)e^{\lambda_n t}+g_b\sum_n\frac{\partial^2f_n(z)}{\partial z\partial z}e^{\lambda_nt}\\\nonumber
&-\bigg(2g_b\frac{\partial V(z,t)}{\partial z}\sum_n\frac{\partial f_n(z)}{\partial z}e^{\lambda_nt}+g^2_b\sum_{m,n}\frac{\partial f_n(z)}{\partial z}\frac{\partial f_m(z)}{\partial z}e^{(\lambda_n+\lambda_m)t}\bigg)\\\nonumber
&-d_{\phi}\bigg(z\frac{\partial V(z,t)}{\partial z}+\epsilon\sum_n\frac{f_n(z)}{\partial z}e^{\lambda_n t}\bigg)+d\bigg(V(z,t)+g_b\sum_nf_n(z)e^{\lambda_n t}\bigg)\\\nonumber
&0=-g_b\sum_n\lambda_nf_n(z)e^{\lambda_n t}+g_b\sum_n\frac{\partial^2f_n(z)}{\partial z\partial z}e^{\lambda_nt}-\bigg(2g_b\frac{\partial V(z,t)}{\partial z}\sum_n\frac{\partial f_n(z)}{\partial z}e^{\lambda_nt}+g^2_b\sum_{m,n}\frac{\partial f_n(z)}{\partial z}\frac{\partial f_m(z)}{\partial z}e^{(\lambda_n+\lambda_m)t}\bigg)\\\nonumber
&-d_{\phi}\bigg(z\frac{\partial V(z,t)}{\partial z}+g_b z\sum_n\frac{f_n(z)}{\partial z}e^{\lambda_n t}\bigg)+d\bigg(V(z,t)+g_b\sum_nf_n(z)e^{\lambda_n t}\bigg),
\end{align}

\noindent then at order $\mathcal{O}(g_b)$ we get the following differential equation,

\begin{align}
\label{epsilon-ex}
e^{\lambda t}\frac{\partial^2f(z)}{\partial z\partial z}=e^{\lambda t}\bigg((\lambda-d)f(z)+d_{\phi}z\frac{\partial f(z)}{\partial z}+2\frac{\partial V(z,t)}{\partial z}\frac{\partial f(z)}{\partial z}\bigg),
\end{align}

\noindent and expanding the last term of the RHS as follows,

\begin{align}
\label{boson_expansion}
e^{\lambda t}\frac{\partial V(z,t)}{\partial z}\frac{\partial f(z)}{\partial z}=&-e^{\lambda t}\frac{\partial \log(U(z,t))}{\partial z}\\\nonumber
=&-e^{\lambda t}\frac{1}{U}\frac{\partial U}{\partial z}\frac{\partial f(z)}{\partial z}\\\nonumber
=&\frac{1}{2}\frac{z}{\lambda t}e^{\lambda t}\frac{\partial f(z)}{\partial z}\\\nonumber
=&\frac{1}{2}\frac{z}{\lambda t}\bigg(1+\lambda t+\frac{\lambda^2t^2}{2!}+\cdots\bigg)\frac{\partial f(z)}{\partial z}\\\nonumber
=&\frac{z}{2}\bigg(\frac{1}{\lambda t}+\frac{\lambda}{\lambda}+\frac{\lambda^2t}{2!\lambda}+\cdots\bigg)\frac{\partial f(z)}{\partial z}\\\nonumber
=&\frac{z}{2}\bigg(\frac{1}{\lambda t}+1+\frac{\lambda t}{2!}+\cdots\bigg)\frac{\partial f(z)}{\partial z},
\end{align}
\noindent then collecting the zeroth order ($\mathcal{O}(t^0)$) terms in expansion parameter $t$ from equation \eqref{epsilon-ex} we get the following differential equation below,

\begin{align}
\frac{\partial^2f(z)}{\partial z\partial z}=&(\lambda-d)f(z)+d_{\phi}z\frac{\partial f(z)}{\partial z}+2\frac{\lambda}{\lambda} \frac{z}{2}\frac{\partial f(z)}{\partial z}\\\nonumber
=&(\lambda-d)f(z)+\big(d_{\phi}+1\big)z\frac{\partial f(z)}{\partial z}\\\nonumber
=&(\lambda-d)f(z)+c_{\phi\lambda}z\frac{\partial f(z)}{\partial z},
\end{align}

\noindent where $c_{\phi\lambda}=d_{\phi}+1$.
\end{widetext}

\nocite{*}
\bibliography{ref}

\end{document}